\documentclass[longauth]{aa} % for the long lists of affiliations 
\usepackage{amsmath}
\usepackage{booktabs, makecell, tabularx}
\usepackage{lipsum}
\setcellgapes{2pt}
\newcolumntype{L}{>{\raggedright\arraybackslash\linespread{0.84}\selectfont}X}

\newcommand{\cii}{\mbox{[C{\small II}]}}

\usepackage{graphicx}
\usepackage[svgnames]{xcolor}
\usepackage{txfonts}
\begin{document}

   \title{The ALMA-CRISTAL survey: Dust temperature and physical conditions of the interstellar medium in a typical galaxy at $z=5.66$}

  \abstract{We present new $\lambda_{\rm rest}=77$ $\mu$m dust continuum observations from the Atacama Large Millimeter/submillimeter Array of HZ10 (CRISTAL-22), a dusty main-sequence galaxy at $z$=5.66 as part of the \cii\, Resolved Ism in STar-forming Alma Large program, CRISTAL.  The high angular resolution of the ALMA Band 7 and new Band 9 data ($\sim{0}''.4$) reveals the complex structure of HZ10, which comprises two main components (HZ10-C and HZ10-W) and a bridge-like dusty emission between them (the Bridge). Using a modified blackbody function to model the dust spectral energy distribution (SED), we constrain the physical conditions of the interstellar medium (ISM) and its variations among the different components identified in HZ10. We find that HZ10-W (the more UV-obscured component) has an SED dust temperature of $T_{\rm SED}$$\sim$51.2$\pm13.1$ K; this is found to be $\sim$5 K higher, although statistically insignificant (less than 1$\sigma$) than that of the central component and previous global estimations for HZ10. Our new ALMA data allow us to reduce by a factor of $\sim$2.3 the uncertainties of global $T_{\rm SED}$ measurements compared to previous studies. HZ10 components have \cii-to-far-infrared (FIR) luminosity ratios and FIR surface densities values consistent with those for local starburst galaxies. However, HZ10-W shows a lower [CII]/FIR ratio compared to the other two components (although still within the uncertainties), which may suggest a harder radiation field destroying polycyclic aromatic hydrocarbon associated with \cii\, emission (e.g., active galactic nuclei or young stellar populations). While HZ10-C appears to follow the tight IRX-$\beta_{\rm UV}$ relation seen in local UV-selected starburst galaxies and high-$z$ star-forming galaxies, we find that both HZ10-W and the Bridge depart from this relation and are well described by dust-screen models with holes in front of a hard UV radiation field. This suggests that the UV emission (likely from young stellar populations) is strongly attenuated in the more dusty components of the HZ10 system.}

   \date{Received September 13th, 2024; accepted October XX, 2024}

	 
    \titlerunning{Physical conditions of ISM at $z=5.67$} 

   %\subtitle{I. Overviewing the $\kappa$-mechanism}

   \author{V. Villanueva
          \inst{1}\fnmsep\thanks{ALMA-ANID Postdoctoral Fellow}
          \and
          R. Herrera-Camus\inst{1}
          \and
          J. Gonz\'alez-L\'opez\inst{2,28}
          \and
          M. Aravena\inst{3}
          \and
          R. J. Assef\inst{3}
          \and
          Mauricio Baeza-Garay\inst{1}
          \and
          L. Barcos-Mu\~noz\inst{4}          
          \and
          S. Bovino\inst{5,6}
          \and
          R. A. A. Bowler\inst{7}
          \and 
          E. da Cunha\inst{8} 
          \and
          I. De Looze\inst{9,10}
          \and
          T. Diaz-Santos\inst{11,12}
          \and
          A. Ferrara\inst{28}
          \and
          N.M. F\"orster Schreiber\inst{13}
          \and
          H. Algera\inst{14}
          \and
          R. Ikeda\inst{15,17}
          \and 
          M. Killi\inst{3}
          \and
          I. Mitsuhashi\inst{16,17}
          \and
          T. Naab\inst{18}  
         \and
          M. Relano\inst{19,20}
          \and
          J. Spilker\inst{21}
          \and 
          M. Solimano\inst{3}
          \and
          M. Palla\inst{22,23}
          \and 
          S.H. Price\inst{24}
          \and
          A. Posses\inst{3}          
          \and
          K. Tadaki\inst{25}
          \and
          K. Telikova\inst{3}
          \and
          H. \"Ubler\inst{26,27}
          }

	 \institute{Departamento de Astronom\'ia, Universidad de Concepci\'on, Barrio Universitario, Concepci\'on, Chile\\
              email{: vvillanueva@astro-udec.cl}
         \and
        Instituto de Astrof\'isica, Facultad de F\'isica, Pontificia Universidad Cat\'olica de Chile, Santiago 7820436, Chile
        \and 
        Instituto de Estudios Astrof\'isicos, Facultad de Ingenier\'ia y Ciencias, Universidad Diego Portales, Av. Ej\'ercito 441, Santiago 8370191, Chile
        \and
        National Radio Astronomy Observatory, 520 Edgemont Road, Charlottesville, VA 22903, USA
        \and
        INAF–Istituto di Radioastronomia, Via Gobetti 101, 40129 Bologna, Italy
        \and
        Chemistry Department, Sapienza University of Rome, P.le A. Moro, 00185 Rome, Italy
        \and
        Jodrell Bank Centre for Astrophysics, Department of Physics and Astronomy, School of Natural Sciences, The University of Manchester, Manchester, M13 9PL, UK
        \and
        International Centre for Radio Astronomy Research, ICRAR M468, 35 Stirling Hwy, Crawley 6009, Western Australia
        \and
        Sterrenkundig Observatorium, Ghent University, Krijgslaan 281 – S9, B9000 Ghent, Belgium
        \and
        Department of Physics \& Astronomy, University College London, Gower Street, London WC1E 6BT, UK
        \and
        Institute of Astrophysics, Foundation for Research and Technology–Hellas (FORTH), Heraklion, GR-70013, Greece
        \and
        School of Sciences, European University Cyprus, Diogenes street, Engomi, 1516 Nicosia, Cyprus
        \and
        Max-Planck-Institut fuer extraterrestrische Physik, Giessenbachstrasse 1, 85748 Garching, Germany
        \and
        Hiroshima Astrophysical Science Center, Hiroshima University, 1-3-1 Kagamiyama, Higashi-Hiroshima, 739-8526 Hiroshima, Japan
        \and 
        Department of Astronomy, School of Science, SOKENDAI (The Graduate University for Advanced Studies), 2-21-1 Osawa, Mitaka, Tokyo 181-8588, Japan
         \and 
        Department of Astronomy, The University of Tokyo, 7-3-1 Hongo, Bunkyo, Tokyo 113-0033, Japan
        \and
        National Astronomical Observatory of Japan, 2-21-1 Osawa, Mitaka, Tokyo 181-8588, Japan 
        \and
        Max Planck Institut f\"ur Astrophysik, Karl-Schwarzschild-Str. 1, D-85741 Garching, Germany
        \and
        Departamento F\'isica Te\'orica y del Cosmos, Universidad de Granada, E-18071 Granada, Spain 
        \and
        Instituto Universitario Carlos I de F\'isica Te\'orica y Computacional, Universidad de Granada, E-18071 Granada, Spain
        \and
        Department of Physics and Astronomy and George P. and Cynthia Woods Mitchell Institute for Fundamental Physics and Astronomy, Texas A\&M University, 4242 TAMU, College Station, TX 77843-4242, USA 
         \and
        Dipartimento di Fisica e Astronomia, Universita di Bologna, via Gobetti 93/2, I-40129 Bologna, Italy
        \and
        INAF - OAS, Osservatorio di Astrofisica e Scienza dello Spazio di Bologna, via Gobetti 93/3, I-40129 Bologna, Italy
        \and
        Department of Physics and Astronomy and PITT PACC, University of Pittsburgh, Pittsburgh, PA 15260, USA 
        \and
        Faculty of Engineering, Hokkai-Gakuen University, Toyohira-ku, Sapporo 062-8605, Japan 
          \and
        Kavli Institute for Cosmology, University of Cambridge, Madingley Road, Cambridge CB3 0HA, UK
        \and
        Cavendish Laboratory, University of Cambridge, 19 JJ Thomson Avenue, Cambridge CB3 0HE, UK 
        \and
        Scuola Normale Superiore, Piazza dei Cavalieri 7, I- 50126 Pisa, Italy 
        \and
        Las Campanas Observatory, Carnegie Institution of Washington, Casilla 601, La Serena, Chile;                                            
              }

	 \maketitle

% \abstract{}{}{}{}{} 
% 5 {} token are mandatory
  \abstract{We present new $\lambda_{\rm rest}=77$ $\mu$m dust continuum observations from the Atacama Large Millimeter/submillimeter Array of HZ10 (CRISTAL-22), a dusty main-sequence galaxy at $z$=5.66 as part of the \cii\, Resolved Ism in STar-forming Alma Large program, CRISTAL.  The high angular resolution of the ALMA Band 7 and new Band 9 data ($\sim{0}''.4$) reveals the complex structure of HZ10, which comprises two main components (HZ10-C and HZ10-W) and a bridge-like dusty emission between them (the Bridge). Using a modified blackbody function to model the dust spectral energy distribution (SED), we constrain the physical conditions of the interstellar medium (ISM) and its variations among the different components identified in HZ10. Our new ALMA data allow us to reduce by a factor of $\sim$2.3 the uncertainties of global $T_{\rm SED}$ measurements compared to previous studies. We find that HZ10-W (the more UV-obscured component) has an SED dust temperature of $T_{\rm SED}$$\sim$51.5$\pm13.1$ K; this is $\sim$5 K higher (although still consistent) than that of the other two components and previous global estimations for HZ10. HZ10 components have \cii-to-far-infrared (FIR) luminosity ratios and FIR surface densities values consistent with those for local starburst galaxies. However, HZ10-W shows a lower [CII]/FIR ratio compared to the other two components (although still within the uncertainties), which may suggest a harder radiation field destroying polycyclic aromatic hydrocarbon associated with \cii\, emission (e.g., active galactic nuclei or young stellar populations). While HZ10-C appears to follow the tight IRX-$\beta_{\rm UV}$ relation seen in local UV-selected starburst galaxies and high-$z$ star-forming galaxies, we find that both HZ10-W and the Bridge depart from this relation and are well described by dust-screen models with holes in front of a hard UV radiation field. This suggests that the UV emission (likely from young stellar populations) is strongly attenuated in the more dusty components of the HZ10 system.}


%
%-------------------------------------------------------------------

\section{Introduction}
\label{sec:intro}

{Accurate measurements of dust temperatures ($T_{\rm dust}$) in star-forming main sequence galaxies are critical to determine their infrared luminosities ($L_{\rm IR}$), star formation rates (SFR), and dust attenuation properties (e.g., the excess of infrared emission compared to the ultra-violet, UV, or Balmer decrement, among others), which are fundamental quantities in the context of galaxy evolution. However, precise estimations of $T_{\rm dust}$ require the detection of the dust continuum emission from both sides of the peak of the infrared spectral energy distribution (SED; e.g., \citealp{Hodge&daCunha2020,daCunha2021}). Although this has been partially achieved at low- and high-redshift galaxies (e.g., \citealt{Bakx2021,Witstok2022,Akins2022,Tsukui2023,Algera2024}), we still need better constraints around the peak of the dust SED in individual main-sequence or typical star-forming main-sequence galaxies at $z \gtrsim 4$.}

The advent of the new generation telescopes (e.g., the Karl G. Jansky Very Large Array, VLA; the NOrthern Extended Millimetre Array, NOEMA; Atacama Large Millimeter/submillimeter Array, ALMA, James Webb Space Telescope, JWST; among others) have revolutionized the exploration of the physical properties of dust, the extragalactic cold neutral/molecular gas, and their close relation with the star formation activity. Nevertheless, we still lack a consensus about the disagreement between $T_{\rm dust}$ estimates at low ($z\lesssim 1-2$) and high ($z\gtrsim 4$) redshift. Although dust peak temperature derived from SED fitting (i.e., the temperature at the wavelength where the SED peaks, $T_{\rm peak}$) in low-$z$ galaxies show a good agreement with predictions from models, estimations in star-forming galaxies at high-$z$ find very low $T_{\rm peak}$ (e.g., compared to predictions by \citealt{Viero2022} at low-$z$). The latter suggests that the physical conditions of dust at high redshift may differ significantly compared to those in the local Universe. In addition, while dust temperatures are well constrained by SEDs densely sampled in wavelength in local galaxies (e.g., \citealt{Villanueva2017,Herrera-Camus2018b}), high-$z$ galaxies usually lack a proper SED coverage (e.g., REBELS, \citealt{Inami2022}; SERENADE, \citealt{Mitsuhashi2023a}), which translates a $T_{\rm dust}$ susceptible to severe biases (e.g., \citealp{Bakx2021,Algera2024}). This can produce significant differences in the derived far IR luminosities ($L_{\rm FIR}$, up to 3 times and more; \citealt{Bouwens2020}), which not only affects our understanding of the dust physical properties in high-$z$ galaxies, but also for the interpretation of their interstellar medium (ISM) and star formation properties (e.g., \citealt{Faisst2020,Herrera-Camus2021}).

Galaxy surveys on the cold neutral gas deepened our knowledge of the star-formation main-sequence (MS) in the local Universe (e.g., \citealt{Brinchmann2004,Whitaker2012, Cano-Diaz2016, Saintonge2016, Colombo2020, Villanueva2024}), and at high-$z$ ($z\gtrsim4$; e.g., \citealt{Capak2015}, and the Alma Large Program to INvestigate C+ at Early times, ALPINE; \citealt{LeFevre2020}). For instance, spectral studies of \cii\, data in high-$z$ galaxies have revealed signatures of high speed outflows ($\sim$400-500 km s$^{-1}$) with mass-outflows rates comparable to their SFRs \citep[e.g.,][]{Gallerani2018, Ginolfi2020a,Herrera-Camus2021}, probably associated to diffuse and extended \cii\, components around galaxies (or \cii\, halos; e.g., \citealt{Fujimoto2019,Fujimoto2020,Fudamoto2022, Pizzati2020,Solimano2024}). Moreover, while low-$z$, normal star-forming galaxies show a tight correlation between the \cii\, luminosity and the SFR surface densities (the $\Sigma_{\rm [CII]}$-$\Sigma_{\rm SFR}$ relation; e.g., \citealt{DeLooze2014,Herrera-Camus2015,Lupi2019}), more extreme star-forming systems depart from the $L_{\rm [CII]}/L_{\rm IR}-L_{\rm IR}$ relation due to a deficit in their \cii\, content. At low-$z$ (e.g., $z\lesssim 0.2 $), these \cii-deficient galaxies are typically dusty, dense starbursts characterized by their hard radiation fields (see \citealt{Malhotra2001,Gracia-Carpio2011,DiazSantos2017,Herrera-Camus2018b} and reference therein). However, the modest spatial resolution achieved by most high-$z$ galaxy surveys (typically $\sim$5-10 kpc at best) does not allow us to verify this effect at the relevant physical scales ($\sim$1 kpc; e.g., \citealt{Shibuya2015}). A detailed characterization of the ISM in high-$z$ galaxies at physical scales comparable to those at low redshift is therefore crucial for a more comprehensive understanding of the nature of this effect.

%The interpretation of the results previously mentioned is severely hampered by the incompleteness of dust SEDs of high-$z$ galaxies that typically includes only one measurement. 
Given that the shape of the dust SED is very sensitive to the temperature (i.e., $T_{\rm SED}$, which refers to the estimation of the $T_{\rm dust}$ from SED modeling), adopting an average value derived from local galaxies can potentially underestimate the IR luminosity by a factor of 5 (e.g., \citealt{Faisst2017}) or even higher (e.g., \citealp{Hodge&daCunha2020}). The observational evidence suggests that the global dust temperature of galaxies are on average warmer at high-$z$ (e.g., \citealt{Magdis2012,Magnelli2014,Bethermin2015,Ferrara2017,Schreiber2018,Liang2019,Sommovigo2020}), which could be the result of either higher star formation activity, obscured AGN (which could provides a substantial part of the emitted power from the galaxy), and/or lower metal content compared to those at low $z$. More accurate constraints on $T_{\rm SED}$ estimations are thus necessary for a better characterization of the IR SEDs and the derivation of the physical conditions of the ISM, and particularly for the dust, in galaxies at $z>4$. 

In order to check how ISM properties (including dust heating) vary on spatially resolved scales in high-$z$ galaxies, we present a systematic analysis of the dust and \cii\, in HZ10, a main-sequence galaxy at $z\approx 5.66$ (see Table \ref{table_1} for more details), as part of the \cii\, Resolved Ism in STar-forming galaxies with ALma survey, CRISTAL (\citealt{Mitsuhashi2023,Solimano2024,Posses2024}; Herrera-Camus et al. in prep.). Based on new ALMA Band 9+7 data (77 and 158 $\mu$m dust continuum) and earlier ALMA Band 8+6 (110 and 198 $\mu$m dust continuum) observations, we cover the region close to the peak of the dust SED.  This allows us to obtain $T_{\rm SED}$ measurements for HZ10 at kpc-scales, significantly improving the accuracy from the previous studies (derivations). The paper is organized as follows: Section \ref{S2_Observations} presents the main features of HZ10, data processing and the ancillary data. In Section \ref{S3_Methods} we explain the methods applied to analyze the data and the equations used to derive the key physical quantities. Finally, in Section \ref{S4_Results} we present our results and discussion, and in Section \ref{S5_Conclusions} we summarize the main conclusions. Throughout this work, we assume a $\Lambda$CDM cosmology, adopting the values $\Omega_{\Lambda}=0.7$, $\Omega_{\rm M}=0.3$ and H$_{\rm o}=70$ km s$^{-1}$ Mpc$^{-1}$, thus resolving physical scales $\approx$ 6.02 kpc per arcsec.

\begin{table}
%\hspace{-1cm}
\resizebox{1.\linewidth}{!}{\begin{tabular}{l l}% <---
\hline
    %\toprule
{Property}  &  { Value}  \\
\hline
(1) $z$   &  {\small 5.659} \\
(2) $D_{\rm L}$ [Mpc] & {\small 54988}\\
(3) Angular Scale Conversion [kpc/${}''$] & {\small 6.023}\\
%(4) $\beta_{\rm d}$   &  {\small 2.15$^{+0.41}_{-0.54}$} \\
(4) log[SFR$_{\rm IR}$/(M$_\odot$ yr$^{-1}$)]   &  {\small 2.48$^{+0.15}_{-0.25}$} \\
(5) log[SFR$_{\rm UV}$/(M$_\odot$ yr$^{-1}$)]   &  {\small1.56$^{+0.06}_{-0.06}$} \\
(6) log[$M_{\star}$/M$_\odot$]   &  {\small 10.39$^{ +0.17}_{-0.17}$} \\
\hline
\hline
\end{tabular}}

\caption{Compilation of the previous-best global physical quantities of HZ10. Row (1): Ly$\alpha$ redshift. Rows (2) and (3): Luminosity distance and the angular scale conversion factor, respectively.% assuming a $\Lambda$CDM cosmology and adopting the values $\Omega_{\Lambda}=0.7$, $\Omega_{\rm DM}=0.3$, and H$_{\rm o}=69.7$ km s$^{-1}$ Mpc$^{-1}$. 
%\noindent Row (4): dust emissivity index ($\beta_{\rm d}$). 
\noindent Row (4): IR star-formation rate (SFR$_{\rm IR}$). Row (5): rest-UV SFR derived from rest-UV luminosity. Row (6): stellar mass derived from SED fitting to the photometry available in the COSMOS or GOODS-S fields using {\tt LE\_PHARE}. Rows (1), (5) and (6) are taken from \cite{Capak2015}. Row (4) are drawn from \cite{Faisst2020}.%,  using the relations given in \cite{Kennicutt1998}.
} 
\label{table_1}
\end{table}

\begin{figure*}
\hspace{-0.3cm}
\includegraphics[width=19cm]{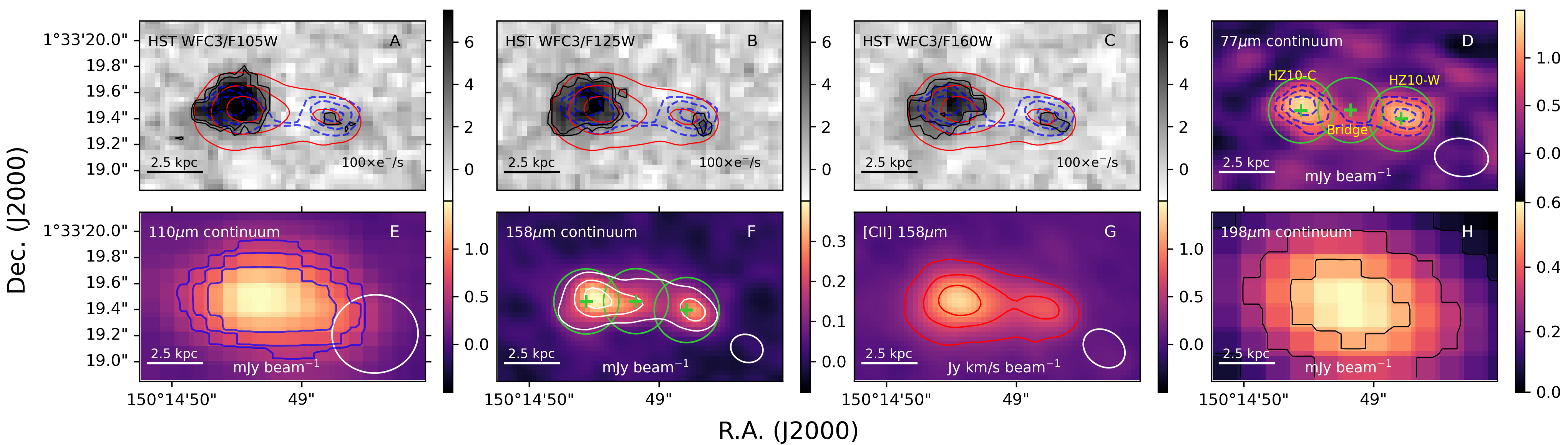}
\vspace{-0.2cm}
\caption{HZ10 HST, ALMA Band 9, 8, and 7, \cii, and Band 6 morphologies in cutouts of ${1.3}'' \times {2.2}''$. {Panels A, B, and C} contain the WFC3 F105W, F125W, and F160W images, respectively, in grey scale. The black contours in the three panels are the [0.9, 1.2, 2.1]$\times 10^{-21}$ erg s$^{-1}$ cm$^{-2}$ $\rm \AA ^{-1}$ levels of the F105W, F125W, and F160W filters. The blue-dashed and red-solid contours correspond to [3$\sigma$, 4$\sigma$, 5$\sigma$] and [6$\sigma$, 12$\sigma$, 18$\sigma$] levels for the 77 $\mu$m continuum and \cii \, integrated intensity, respectively. {From D to H:} Panels include the 77 $\mu$m ALMA continuum, 110 $\mu$m ALMA continuum, 158 $\mu$m ALMA continuum, and 158 $\mu$m \cii, and 198 $\mu$m ALMA continuum images, respectively. Contours levels in panels D and G are the same as in panel A, B, and C, for 77 $\mu$m continuum, and \cii \, integrated intensity, respectively. While blue-solid contours in panels E and H [3$\sigma$, 4$\sigma$, 5$\sigma$] levels for 110 $\mu$m and 198 $\mu$m ALMA continuum maps, respectively, the white-solid contours in panel G are the [6$\sigma$, 4$\sigma$, 5$\sigma$] levels for the 158 $\mu$m ALMA continuum maps. Finally, while green crosses in panel B are the centers of the Se\'ersic profiles computed for the three sources (e.g., HZ10-C, HZ10-W, and the Bridge) analysed in this work (see \S\ref{Parametric_morphology}), the green circles correspond to the apertures used to perform the angular resolved analysis (see \S\ref{Dust_temperature} for more details).} 
\label{fig_1}
\end{figure*}

\section{Observations}
\label{S2_Observations}
%{\color{magenta} POR QUE NO OBSERVACIONES?}

\subsection{The ALMA-CRISTAL sample}
\label{cristal_sample}

CRISTAL is an ALMA Cycle-8 large program (2021.1.00280.L; PI: R. Herrera-Camus), aiming to get spatially resolved \cii\, line and 870 $\mu$m dust-continuum emission data for star-forming main-sequence galaxies at $z\sim4-6$ (Herrera-Camus et al. in prep.). 
%The CRISTAL sample selection and data reduction are described in more detail in other CRISTAL studies (e.g., \citealt{Mitsuhashi2023,Solimano2024}; Posses et al. in prep.; Herrera-Camus et al. in prep.); here we summarize the main features.

The CRISTAL sample is drawn from \cii\, detected galaxies from the ALPINE survey \citep{LeFevre2020}, all located in the COSMOS or GOODS-S fields. Galaxies were selected based on spectral energy distribution (SED) modelling to have: i) specific star-formation rates (sSFR) within a factor of three from the MS; ii) ancillary HST/WFC3 rest-frame UV data; and iii) stellar masses log$[M_{\star}/$M$_{\odot}]\geq 9.5$. Additionally, six extra galaxies from the COSMOS field (which meet the selection criteria mentioned above) were added: HZ04, HZ07, HZ10, DC818760, DC873756, VC8326, all with comparable spatial resolutions and sensitivities to those of the main sample (from ALMA programs 2018.1.01359.S and 2019.1.01075.S; PI: Manuel Aravena, 2018.1.01605.S; PI: Rodrigo Herrera-Camus, 2019.1.00226.S; PI: Edo Ibar). The latter also meet the selection criteria mentioned above. %The latter complete a final sample of 25 galaxies.

%\vspace{1.cm}
\subsection{The HZ10 data}
\label{alma_data}

%{\color{magenta} DATA ARE PLURAL, DATUM IS SINGULAR}
HZ10 stands as one of the best CRISTAL sources in order to investigate the physical conditions of the dust at $z=4-6$ due to the rich multi-wavelength data available. The latter come mostly from previous studies primarily oriented to derive the main features of HZ10 using integrated quantities (see Table \ref{table_1} and references therein). To analyze HZ10 (CRISTAL-22 in the CRISTAL survey), we use data from different ALMA cycles/projects listed and described as follows:

\begin{itemize}
    
    \item {\bf New Band 9 data:} Observations of the rest-frame 77 $\mu$m continuum were taken on August 19th, 2022, during Cycle 8 ($\nu_{\rm band9}=682$ GHz; panel D in Fig. \ref{fig_1}) as part of project 2022.1.00678.S (P.I.: R. Herrera-Camus). The integration time was $\sim$77 minutes on-source, achieving a sensitivity of $\sim 0.24$ mJy beam$^{-1}$ over $\Delta \nu=13$ GHz and a beamsize of $\sim {0}''.414 \times {0}''.293$. 

    \item {\bf Band 8:} Observations of the rest-frame 110 $\mu$m continuum were taken on January 9th, 2019, during Cycle 6 ($\nu_{\rm band8}=411.4$ GHz; panel E in Fig. \ref{fig_1}) as part of the project 2018.1.00348.S (P.I.: A. Faisst). The integration time was $\sim48$ minutes on-source, achieving a sensitivity of $\sim 0.07$ mJy beam$^{-1}$ over $\Delta \nu =8$ GHz, and a beamsize of $\sim {0}''.663 \times {0}''.601$.
    
    \item {\bf Band 7:} Observations of the rest-frame \cii\ 158 $\mu$m line emission and the corresponding continuum were taken between March 26th and March 30th, 2021 during Cycle 7 ($\nu_{\rm band7}=285.3$ GHz; panels F and G, respectively, in Fig. \ref{fig_1}) as part of the project 2019.1.01075.S (P.I.: M. Aravena). The integration time on source-was $\sim93$ minutes. While the sensitivity and the beamsize achieved for the 158 $\mu$m continuum are $\sim 1.6 \times 10^{-2}$ mJy beam$^{-1}$ over $\Delta \nu\approx6$ GHz, and $\sim {0.252}'' \times {0.212}''$, respectively, for the \cii\, emission line are $\sim 0.14$ mJy beam$^{-1}$ at $\Delta \nu\approx47$ MHz, and $\sim {0}''.264 \times {0}''.225$, respectively. 

    \item {\bf Band 6:} Observations of the rest-frame 198 $\mu$m continuum were taken on January 5th 2015 during Cycle 3 ($\nu_{\rm band6}=220.5$ GHz; panel H in Fig. \ref{fig_1}) as part of the project 2015.1.00388.S (P.I.: N. Lu). The integration time was $\sim54$ minutes on-source, achieving a sensitivity of $\sim 0.03$ mJy beam$^{-1}$ over $\Delta \nu=8$ GHz, and a beamsize of $\sim {1}''.31 \times {1}''.03$.   
    
\end{itemize}

\noindent HZ10 was included in the CRISTAL sample as part of one of the CRISTAL pilot programs that resulted as a combination of the ALMA programs 2019.1.01075.S (P.I.: M. Aravena), and 2012.1.00523.S (P.I.: P. Capak; see \citealt{Capak2015} for more details); however, the latter was not considered in this work due to its coarser angular resolution compared to the former. The details of the data reduction of the Band 7 observations are presented in \cite{Solimano2024}; among the main features their data processing, the self-calibrated and combined measurement sets were processed with CRISTAL's reduction pipeline as described in Herrera-Camus et al. (in prep.). Briefly, it starts by subtracting the continuum on the visibility space using using the Common Astronomy Software Application ({\tt CASA}; \citealt{CASA2022}) {\tt uvcontsub} task. After that, it runs {\tt tclean} with {\tt automasking} multiple times, producing cubes with different weightings and channel widths. In all cases the data are cleaned down to 1$\sigma$. In this paper, we use datacubes with 20 km s$^{-1}$ channel width and Briggs ({\tt robust=0.5}) weighting. The ancillary ALMA Band 8 and Band 6 observations are presented in \cite{Faisst2020}, using Briggs weighting (\citealt{Briggs1995}) for the image reconstruction with a {\tt robust =0.5}.

%{\color{red} [Jorge: here we need your help to describe the Band 9 data reduction]}

%Herrera-Camus et al. (in prep.) include a detailed description of the calibration and measurements combination procedure for ALMA Band 7 data. In a nutshell, they implement a continuum subtraction on the visibility space using the Common Astronomy Software Application ({\tt CASA}; \citealt{CASA2022}). Next, the pipeline runs the task {\tt tclean} with automasking multiple times, cleaning the data down to a threshold of $1\sigma$ and producing cubes with different weightings. %The final \cii\, 20 km/s channel widths datacubes and continuum. 
%\noindent To achieve a good compromise between synthesized beamsize and signal-to-noise, the procedure uses a Briggs weighting parameter of $=0.5$ (or {\tt robust} parameter; \citealt{Briggs1995}). {\color{red} Here we have to include something specific about the Band 9 data reduction; maybe Jorge Gonzalez can help}. Finally, Band 8 and Band 6 ALMA data are calibrated using {\tt CASA} 5.4.0-70 and 4.5.1 and running the {\tt scriptForPI.py} script with {\tt robust=0.5} and 1.0, respectively. 

In addition, we include HST data for HZ10 as part of the project 13641 (P.I.: Peter Capak) and retrieved from the Barbara A. Mikulski Archive for Space Telescopes (MAST\footnote{\footnotesize https://mast.stsci.edu/portal/Mashup/Clients/Mast/Portal.html}). The data comprise Wide Field Camera 3 images (WFC3), including F105W (panel A in Fig. \ref{fig_1}), F125W, and F160W bands. These bands cover a rest-frame wavelength range between $\sim$1200 to 2200$\rm \AA$.%, which corresponds to UV emission at the redshift of HZ10 (i.e., $z\approx 5.66$).

\section{Methods and products}
\label{S3_Methods}

\subsection{Basic equations and assumptions}
\label{Basic_equations}

To compute the integrated \cii\, line emission and the rest-frame 77 $\mu$m, 110 $\mu$m, 158 $\mu$m, and 198 $\mu$m continuum fluxes, we use the following equation:
\begin{equation}
S_{\rm i}  = \int_{A} I_{\rm i}(r) dA,
\label{eq_0}
\end{equation}

\noindent where $A$ is the area of a circular aperture with a diameter equivalent to the major axis of the 198 $\mu$m continuum beamsize (i.e., the coarsest angular resolution among the full dataset; $\sim {1}''.2$) and centered at the position of the Bridge (see \S\ref{Dust_temperature} for more details). Here, $I_{\rm i}$ is the velocity integrated flux density (in Jy km/s beam$^{-1}$) for \cii, and flux density (in Jy beam$^{-1}$) for dust-continuum bands. Finally, $i=$ \cii, 77$\mu$m, $110\mu$m, $158\mu$m, and $198\mu$m. 

We also use Equation \ref{eq_0} to compute resolved values for the three components of HZ10 identified in this work. In this case, $A$ is the area of a circular aperture with a diameter equal to the major axis of the 77 $\mu$m continuum beamsize (i.e., the coarser angular resolution between ALMA Band 7 and 9 data; $\sim {0}''.4$) and centered at the locations of the three components identified in HZ10 (green circles in panels B and D from Fig.\ref{fig_1}).

We use the SED parametrization from \cite{Casey2012} to compute the best spectral energy distribution (SED) fitting, which corresponds to:% and a mid-IR power law:

\begin{equation}
S_{\lambda}= N_{\rm bb} \times f(\lambda; \beta_{\rm d}, T_{\rm SED}), %+ N_{\rm pl} \lambda^{\alpha} e^{{-(\lambda/\lambda_{\rm c})}^2},
\label{eq_1}
\end{equation}
%with

%\begin{equation}
%N_{\rm pl}= N_{\rm bb} \times f(\lambda_{\rm c}; \beta_{\rm d}, T_{\rm SED})
%\label{eq_2}
%\end{equation}

with 
\begin{equation}
f(\lambda; \beta_{\rm d}, T_{\rm SED}) \equiv \frac{(1-e^{{-(\lambda_0 /\lambda_{\rm c})}^{\beta_{\rm d}}})(\frac{c}{\lambda})^3}{e^{{(hc)/(\lambda k T_{\rm SED} )}}-1},
\label{eq_3}
\end{equation}

\noindent where $N_{\rm bb}$ is the normalization, %$\alpha$ is the slope of the mid-IR power law, 
\noindent $\beta_{\rm d}$ is the emissivity index, $T_{\rm SED}$ is the SED dust temperature, $\lambda_0$ is the wavelength at optical depth $\tau =1$, $\lambda_{\rm c}$ is the power-law turnover wavelength, and $c$ is the speed of light. We use %$\alpha=2.0$ and 
\noindent $\lambda_0 =100$ $\mu$m for the global and resolved HZ10 SED fittings as adopted in \cite{Faisst2020}. From \cite{Casey2012}, we also adopt the parametrization of $\lambda_{\rm c}= \frac{3}{4} L(\alpha,T)$, where $L(\alpha,T_{\rm SED}) = [(b_1 + b_2 \alpha)^{-2} + (b_3 + b_4 \alpha)\times T_{\rm SED}]^{-1}$ (with $b_1 = 26.68$, $b_2 = 6.246$, $b_3 = 1.905 \times 10^{-4}$, and $b_4 = 7.243 \times 10^{-5}$, and where $\alpha=2.0$; see \citealt{Casey2012} for more details). We note that upcoming CRISTAL papers (e.g.,  Li et al. submitted) will use other SED fitting methodologies (e.g., SED fitting codes based on the implementation of energy balances such as {\tt MAGPHYS}; \citealt{daCunha2008}) to provide independent estimations of both dust temperature and its related physical quantities.

We compute the peak dust temperature ($T_{\rm peak}$; e.g. \citealt{Bethermin2015,Schreiber2018}), which is derived from the IR emission by the Wien's displacement law,
\begin{equation}
T_{\rm peak} = \frac{2.898 \times 10^3 ({\rm \mu m \, K})}{\lambda_{\rm peak} ({\rm \mu m})}.
\label{eq_4}
\end{equation}

\noindent Since contribution from background CMB heating could potentially affect dust temperatures at $z>5$ (e.g., \citealt{daCunha2013,Faisst2020}), we apply CMB corrections to our $T_{\rm SED}$ estimations using Equation (12) from \cite{daCunha2013}:

\begin{equation}
T_{\rm dust}(z) = [(T^{z=0}_{\rm dust})^{4+\beta_{\rm d}} + (T^{z=0}_{\rm CMB})^{4+\beta_{\rm d}} ([1+z]^{4+\beta_{\rm d}}-1)  ]^{\frac{1}{4+\beta_{\rm d}}},
\label{eq_5}
\end{equation}
\noindent where $T^{z=0}_{\rm dust}$ and $T^{z=0}_{\rm CMB}$ are the dust temperature and the CMB temperature measured at $z=0$, respectively. Along this paper, we adopt $T^{z=0}_{\rm CMB}=2.73$ K .

To compute the total far-IR luminosity ($L_{\rm FIR}$), we perform a numerical integration of Equation \ref{eq_1} in the wavelength range between 42.5 and 125.5 $\mu$m \citep[e.g.,][]{Helou1988}. To do so, we adopt the best-SED fitting parameters of the source (see \S\ref{Dust_temperature} for more details), and we compute the total far-IR luminosity, $L_{\rm FIR}$, by integrating the flux between 42.5 and 122.5$\mu$m (as described in \citealt{Faisst2020}).  Similarly, we also compute the total IR luminosity ($L_{\rm IR}$) by integrating numerically the best-fitting SED but in the wavelength range between 8 and 1000 $\mu$m.%use the following equation: 

%\begin{equation}
%L_{\rm FIR} = 4\pi D^{2}_{\rm L}(z) \int^{\rm 125.5 \mu m}_{\rm 42.5 \mu m} S_{\nu}(c/\lambda^2) d\lambda.
%\label{eq_5}
%\end{equation}

%\noindent Similarly, we also compute the total IR luminosity ($L_{\rm IR}$) by integrating numerically the best-fitting SED but in the wavelength range between 8 and 1000 $\mu$m.

We obtain the \cii \, luminosity (in $L_{\odot}$) using the following equation \citep{Solomon&VandenBout2005}:

\begin{equation}
L_{\rm [CII]} = 3.25 \times 10^7 S_{\rm {CII}} \Delta v \, \nu^{-2}_{\rm obs} \, D^{2}_{\rm L} \, (1+z)^{-3},
\label{eq_6}
\end{equation}

\noindent where $S_{\rm {CII}} \Delta v$ is the velocity integrated flux (in Jy km/s beam$^{-1}$), $D^{2}_{\rm L}$ is the luminosity distance (in Mpc), $\nu_{\rm obs}$ is the observed frequency (in GHz), and $z$ is the redshift.

The UV spectral slope ($\beta_{\rm UV}$) is derived from HST WFC3 F125W and F160W images and using Equation 1 in \cite{Liang2021}:

\begin{equation}
\beta_{\rm UV} = \frac{\log(f_{\lambda, 0.19})-\log(f_{\lambda, 0.23})}{\log(\lambda_{0.19})-\log(\lambda_{0.23})},
\label{eq_7}
\end{equation}

\noindent where $f_{\lambda, 0.19}$ and $f_{\lambda, 0.23}$ are the specific fluxes at ${\rm 1877 \AA}$ and ${\rm 2311 \AA}$ rest frame taken from the F125W and F160W images, respectively. We compute $\beta_{\rm UV}$ by taking advantage of its almost constant value along the wavelength range ${\rm 1260 < \lambda < 3200 \AA}$, avoiding contamination by the ${\rm 2175 \AA}$ ``bump'' feature.

Finally, we compute the IR excess (IRX) using the equation (e.g., \citealt{Meurer1999,Popping2017})
\begin{equation}
{\rm IRX} = L_{\rm IR}/L_{\rm UV},
\label{eq_8}
\end{equation}

\noindent where $L_{\rm UV}$ is the monochromatic rest-frame UV luminosity at ${\rm 1600\AA}$. In order to derive $L_{\rm 1600\AA}$, we use the HST WFC3/F105W image (see panel A in Fig. \ref{fig_1}) and the {\tt PHOTFLAM} constant to convert the flux from e$^{-}$/s to erg s$^{-1}$/cm$^{2}$/$\rm \AA$.%erg/cm$^{2}$/$\rm \AA$/e$^{-}$.

\subsection{Parametric 2D fitting}
\label{Parametric_morphology}

As Fig. \ref{fig_1} shows, the ALMA Band 9 and 7 high angular resolution observations reveal that HZ10 is a complex system with (at least) two main components: a central one (HZ10-C) and one in the west direction (HZ10-W). To derive the morphological parameters of these two components, we use the 2D light profile modeling code {\tt PYAUTOGALAXY}\footnote{https://github.com/Jammy2211/PyAutoGalaxy} \citep{Nightingale2023}, which is based on {\tt PYAUTOFIT} \citep{Nightingale2021}. Implementing the image-based filter {\tt PYAUTOGALAXY} mode for a faster workflow, we generate the noise-map by feeding the full covariance matrix into the calculation of the likelihood using the code {\tt ESSENCE}\footnote{https://github.com/takafumi291/ESSENCE} \citep{Tsukui2023}. %from the signal free part of ALMA Band 9 and Band 7 datacubes. 
\noindent We model the HZ10-C and HZ10-W morphologies by fitting a single 2D S\'ersic profile \citep{Sersic1968}, with a total of seven free parameters: the coordinates of the source's center (R.A. and Dec.), the coordinate of the vector of the S\'ersic profile's minor and major axes $(x,y)$, the effective radius ($R_{\rm e}$), the S\'ersic's index $(n)$, and the intensity at the center ($I_0[r=0]$). 

We perform three independent fits to derive the morphologies of the rest-frame 77 $\mu$m continuum (Band 9), and the 158 $\mu$m continuum and \cii\, line emission maps (Band 7). While the continuum maps are directly generated after running the {\tt tclean} task in {\tt CASA}, we produce the \cii\, intensity map (or moment 0 maps) by fitting a Gaussian function to the line profile and integrating the emission in the spectral range [$\mu$-FWTM,$\mu$+FWTM] (where $\mu$ is the central frequency and FWTM is the full width at one tenth of maximum of the Gaussian profile). We use the following method to compute the best morphological parameters of the emission:

\begin{enumerate}
    
    \item We estimate the centroid using 2D Gaussian functions (R.A. and Dec.) of the 77 $\mu$m continuum emission of HZ10-C and HZ10-W. We use these coordinates as an initial guess to look for the centers of the two 2D S\'ersic profiles. 
    
    \item Then, we use the task {\tt minimize} from the {\tt PYTHON} package {\tt lmfit} \citep{Newville2015}. We perform a two step best-parameters searching: we adopt the {\tt least\_squares} method, followed by the {\tt emcee} method (the latter looks for the maximum likelihood via a Monte-Carlo Markov Chain). To do so, we constrain the values of $n$ and $I_0$ within the ranges $[0.1,3]$ and $[0,2I_{\rm max}]$, respectively ($I_{\rm max}$ is the maximum value of the intensity map).

    \item Finally, we repeat step 2 for the 158 $\mu$m continuum and the \cii\, line emission maps, using this time the 77 $\mu$m continuum best parameters as a prior (i.e., centroids, S\'ersic indexes, $I_{\rm max}$).        
\end{enumerate}

Table \ref{table_2} lists the best parameters and uncertainties of the fitting procedure. We note that the differences in DEC in Table \ref{table_2} are larger than the spatial resolutions of the observations in each band (see \S\ref{alma_data}).

\section{Results and Discussion}
\label{S4_Results}

\begin{table}%[ht]
\par\rule{\columnwidth}{1.pt}
\makegapedcells

\centering
\setlength\extrarowheight{-6pt}
\begin{tabularx}{1\linewidth}{>{\raggedright}p{8em} L}
Parameter   &   Value    \\
    &    {\footnotesize R.A.:10$^{\rm h}$00$^{\rm m}$, Dec.: +01$^{\circ} {33}'$}    \\
\bottomrule
\end{tabularx}

\begin{tabular}{cccccccccc}
   & & &  & {\it HZ10-C} & & & & {\it HZ10-W} \\
\end{tabular}
\par\rule{\columnwidth}{1.pt}

%\begin{tabular}{c}
%   \\
%\end{tabular}
%\bottomrule

{\it 77 $\mu$m continuum}
\makegapedcells

%\addtolength{\tabcolsep}{-3pt}
\begin{tabular}{l l l}
{\footnotesize Center (R.A.)}  &  {\footnotesize 59.304$^{\rm s} \pm$0.037$^{\rm s}$} & {\footnotesize 59.255$^{\rm s} \pm$0.091$^{\rm s}$}  \\
{\footnotesize Center (Dec.)}   &   {\footnotesize ${19.507}''$$\pm$${0.038}''$} & {\footnotesize ${19.408}''$$\pm$${0.037}''$}\\
{\footnotesize Effective radius $R_{\rm e}$ [kpc]}   &   {\footnotesize 1.26$\pm$0.26} & {\footnotesize 1.43$\pm$0.34}\\
{\footnotesize S\'ersic index ($n_{\rm 77\mu m}$)} & {\footnotesize1.70$\pm$0.42} & {\footnotesize1.84$\pm$0.38}\\ 
{\footnotesize Axis ratio (min/maj)} & {\footnotesize 0.66$\pm$0.10} & {\footnotesize 0.60$\pm$0.08} \\ 
{\footnotesize $I_{0}$(r=0) [mJy/bm]} & {\footnotesize 0.61$\pm$0.25} & {\footnotesize 0.22$\pm$0.24}\\
    \bottomrule
\end{tabular}

{\it 158 $\mu$m continuum} %\label{tab:Question-2}

%\addtolength{\tabcolsep}{-3pt}
\begin{tabular}{l l l} % <---
    %\toprule
{\footnotesize Center (R.A.)}  &   {\footnotesize 59.297$^{\rm s}$$\pm$$0.039^{\rm s}$} &  {\footnotesize 59.254$^{\rm s}$$\pm$$0.040^{\rm s}$}  \\
{\footnotesize Center (Dec.)}   &  {\footnotesize${19.486}''$$\pm$${0.037}''$} & {\footnotesize${19.396}''$$\pm$${0.038}''$} \\
{\footnotesize Effective radius $R_{\rm e}$ [kpc]}   &   {\footnotesize 0.78$\pm$0.25} & {\footnotesize 1.00$\pm$0.23}\\
{\footnotesize S\'ersic index ($n_{\rm 158\mu m}$)} & {\footnotesize 1.63$\pm$0.49} & {\footnotesize 1.54$\pm$0.19}\\ 
{\footnotesize Axis ratio (min/maj)} & {\footnotesize 0.58$\pm$0.07} & {\footnotesize 0.59$\pm$0.04}\\ 
{\footnotesize $I_{0}$(r=0) [mJy/bm]} & {\footnotesize 0.24$\pm$0.04} & {\footnotesize 0.12$\pm$0.03}\\
    \bottomrule
\end{tabular}

{\it \cii \, line emission} %\label{tab:Question-2}
 \makegapedcells

%\addtolength{\tabcolsep}{-3pt}
\begin{tabular}{l l l}% <---
    %\toprule
{\footnotesize Center (R.A.)}  &   {\footnotesize 59.298$^{\rm s} \pm$0.039$^{\rm s}$} & {\footnotesize 59.253$^{\rm s}$$\pm$$0.048^{\rm s}$}  \\
{\footnotesize Center (Dec.)}   &  {\footnotesize ${19.492}''$$\pm$${0.040}''$} &  {\footnotesize ${19.405}''$$\pm$${0.039}''$}\\
{\footnotesize Effective radius $R_{\rm e}$ [kpc]}   &   {\footnotesize 1.49$\pm$0.23} & {\footnotesize 1.03$\pm$0.10}\\
{\footnotesize S\'ersic index ($n_{\rm [CII]}$)} &  {\footnotesize2.76$\pm$0.12} & {\footnotesize2.21$\pm$0.31} \\ 
{\footnotesize Axis ratio (min/maj)} &  {\footnotesize0.64$\pm$0.02} & {\footnotesize0.48$\pm$0.07} \\ 
{\footnotesize $I_{0}$(r=0) [Jy/bm km/s]} & {\footnotesize1.18$\pm$0.09} & {\footnotesize1.30$\pm$0.27}\\
    \bottomrule
    \bottomrule
\end{tabular}
\caption{Results of the parametric 2D fitting of S\'ersic profiles, using {\tt PYAUTOGALAXY}, of the 77 $\mu$m continuum (top), 158 $\mu$m continuum (middle), and \cii\, 158 $\mu$m line emission (bottom) for HZ10-C and HZ10-W.} 
\label{table_2}
\end{table}

\begin{figure*}
\hspace{2.5cm}
\includegraphics[width=13cm]{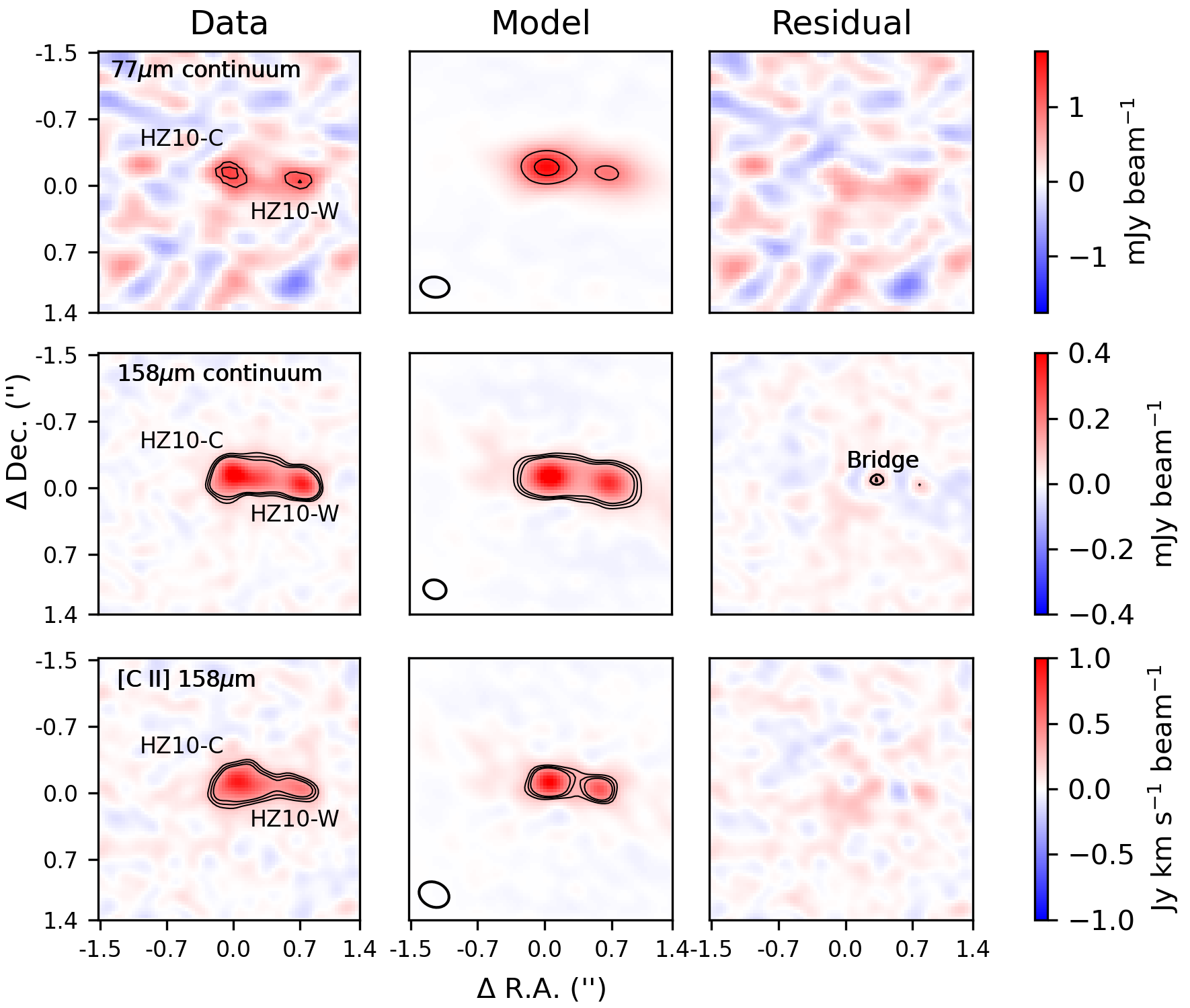}
\caption{Results from the parametric 2D modeling of HZ10 using {\tt PYAUTOGALAXY}, in panels of ${3}'' \times {3}''$. While left panels contain the observed emission (colormap and black-solid contours), middle and right panels show the best model using 2D S\'ersic profiles (see Table \ref{table_2} for more details) and the residuals after subtracting the maximum likelihood model, respectively. From top to bottom, panels contain the rest-frame 77 $\mu$m ALMA continuum (top row), 158 $\mu$m ALMA continuum (middle row), and \cii \, 158 $\mu$m line emission (top row). The beamsize of the data are represented by ellipses at the bottom-left of middle panels. For all the subplots, the contour are the [0 (dashed), 4$\sigma$, 5$\sigma$, and 6$\sigma$] levels. The figure confirms the binary nature of HZ10, which can be decomposed in HZ10-C (at the center), HZ10-W (to the left), and the Bridge in between the two main components (with at least a 5$\sigma$ significance on the residual 158 $\mu$m continuum map). The latter seems to reflect the extended dusty component connecting HZ10-C and HZ10-W.}
\label{fig_2}
\end{figure*}

\subsection{The structure of HZ10}
\label{HZ10_structure}

Figure \ref{fig_2}\, shows the results of the 2D parametric fitting, confirming the complex structure of HZ10. The figure shows this system has (at least) two main components, HZ10-C and HZ10-W, each of them slightly showing different morphologies depending on the datasets. On the one hand, HZ10-W shows a slightly more extended distribution of the \cii\, emission than HZ10-C; on the other hand, the two components have surprisingly similar S\'ersic profiles when analyzing the 77 and 158 $\mu$m continuum maps.  These results may reflect significant dust content (relative to the gas) in HZ10-W that causes severe dust attenuation, as evidenced by its faint UV emission (shown in panel A of Fig. \ref{fig_1}).

\begin{figure}
    %\hspace{-0.5cm}
  \includegraphics[width=9.cm]{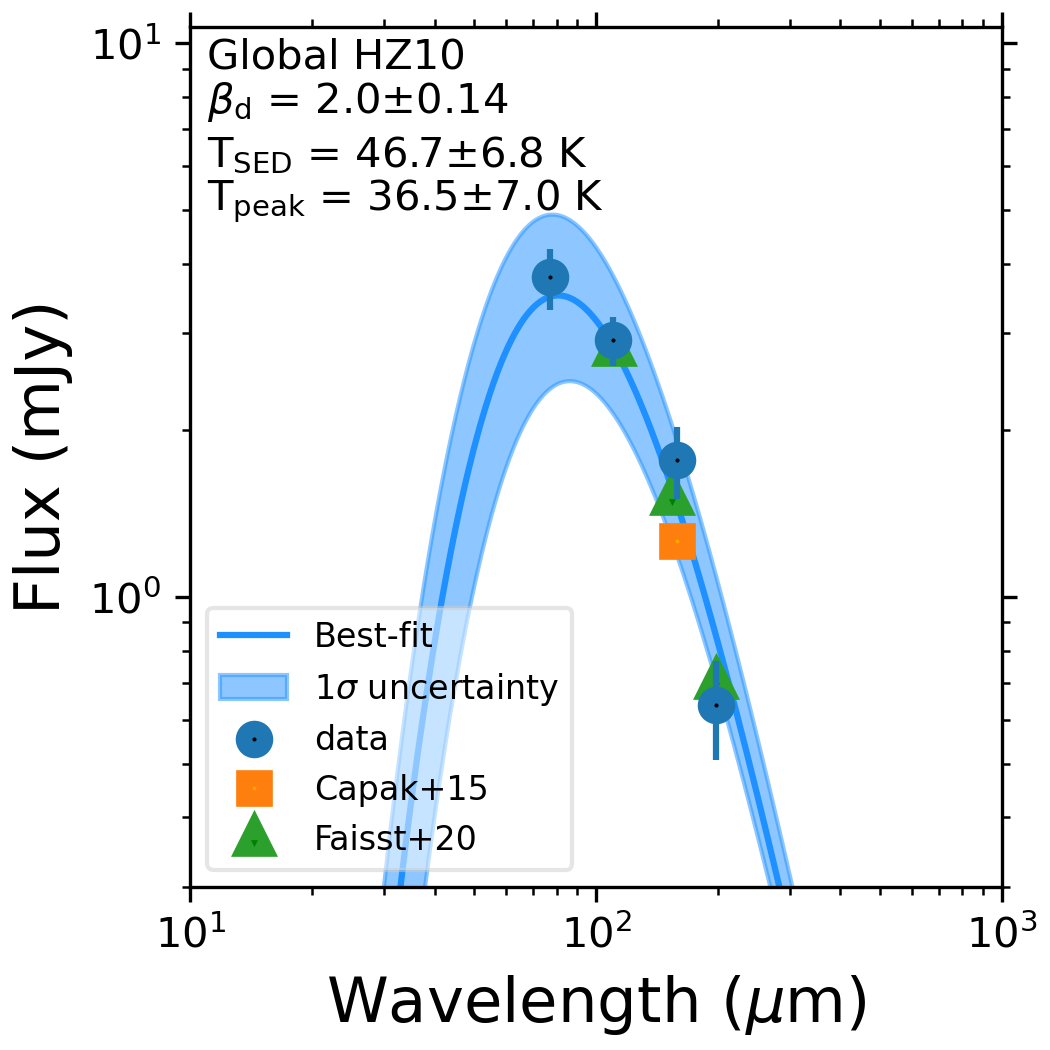} 
  \caption{Best-fitting IR SEDs for the global ALMA Bands 9, 8, 7, and 6 continuum emission of HZ10 (77 $\mu$m, 110 $\mu$m, 158 $\mu$m, and 198 $\mu$m, respectively). The wavelength is given in rest frame. The blue-solid line corresponds to the best-modified SED given by equations \ref{eq_1}-\ref{eq_4} and using the maximum likelihood parameters (see top-left corner). The combination  between new ALMA Band 9 and 7 continuum measurements (77 $\mu$m and 158 $\mu$m rest frame, respectively) allow us to probe the peak of the dust SED, and hence constrain these parameters accurately. In addition, we have included estimations of the global continuum emission flux for HZ10 from \cite{Capak2015} (orange square) and \cite{Faisst2020} (green triangles) to contrast our estimations with previous results.}
  \label{fig_3}
\end{figure}

\begin{figure*}
    %\hspace{-0.3cm}
  \includegraphics[width=6.cm]{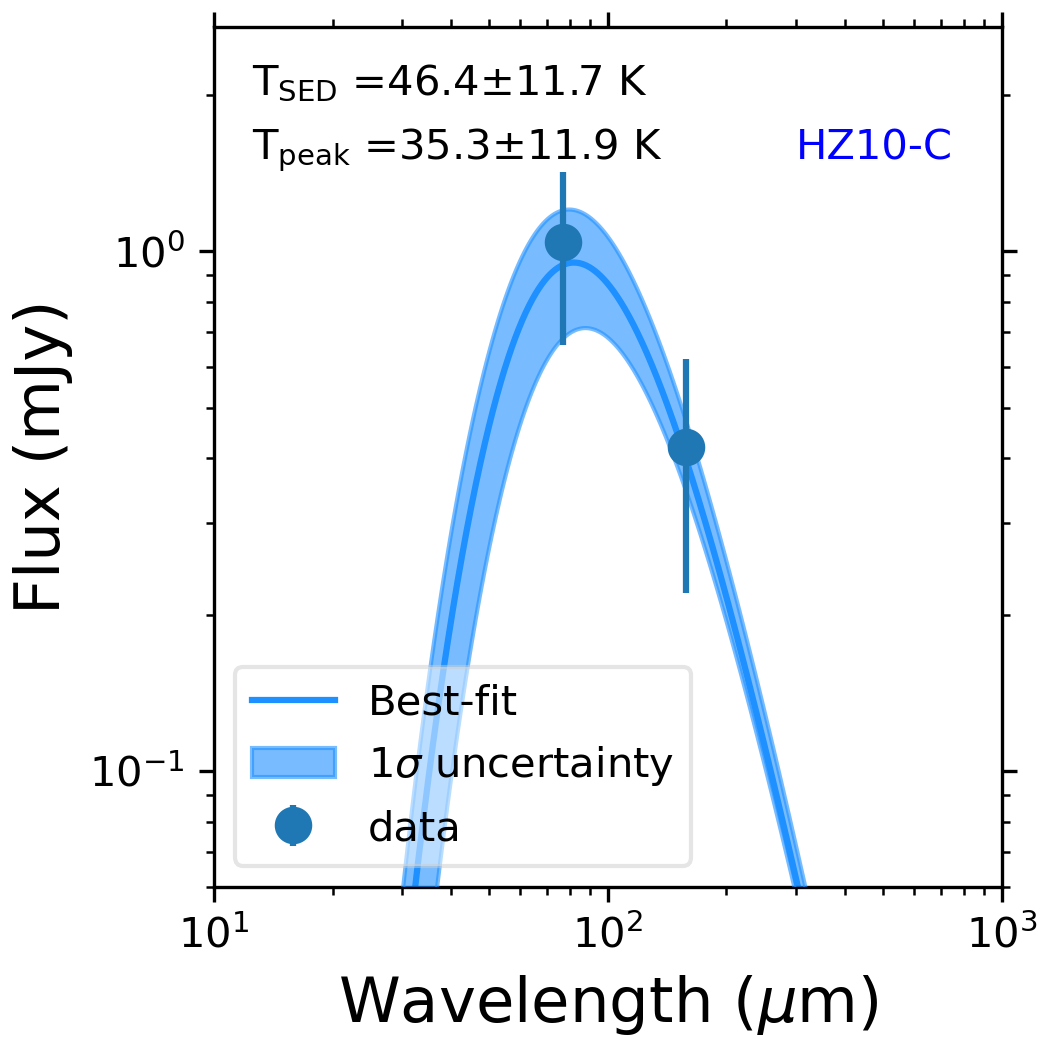} 
  \includegraphics[width=6.cm]{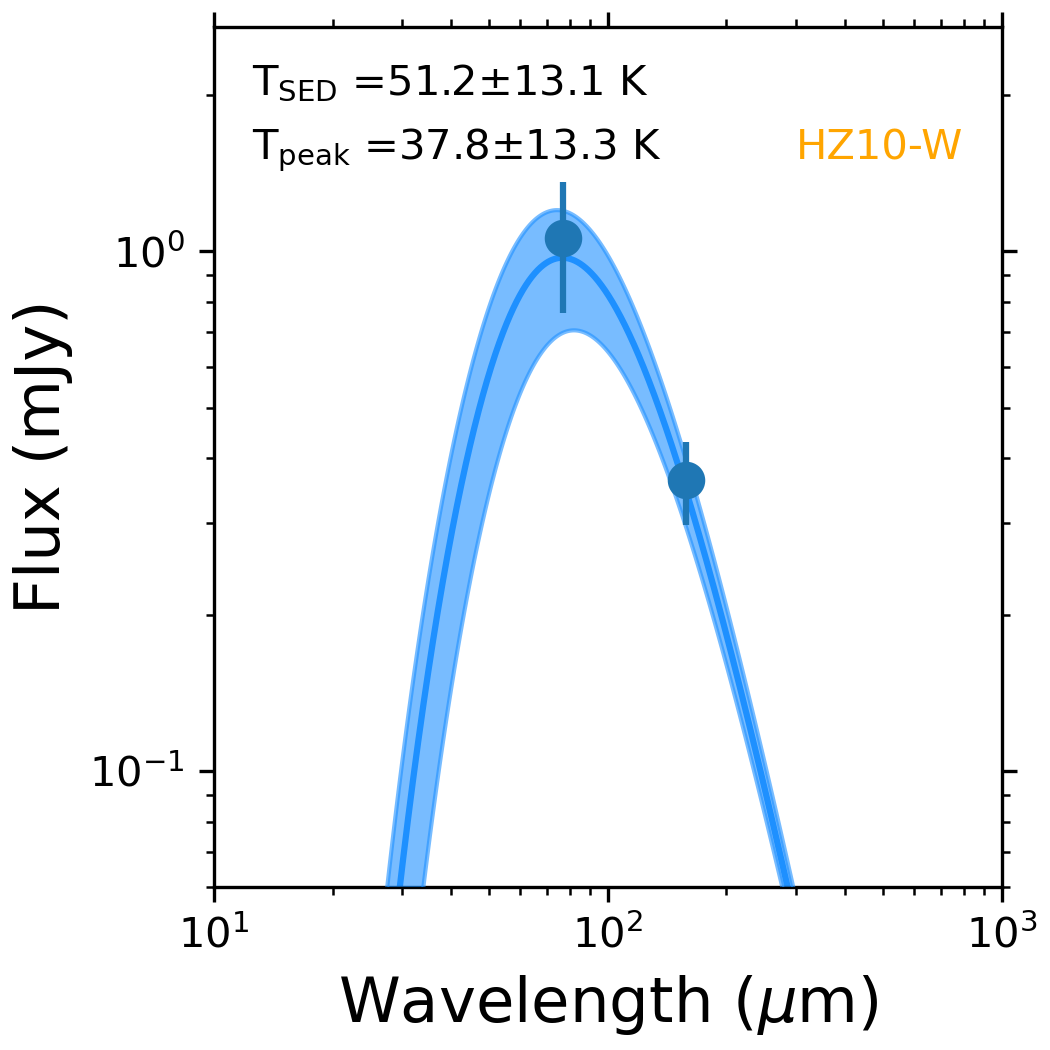}
  \includegraphics[width=6.cm]{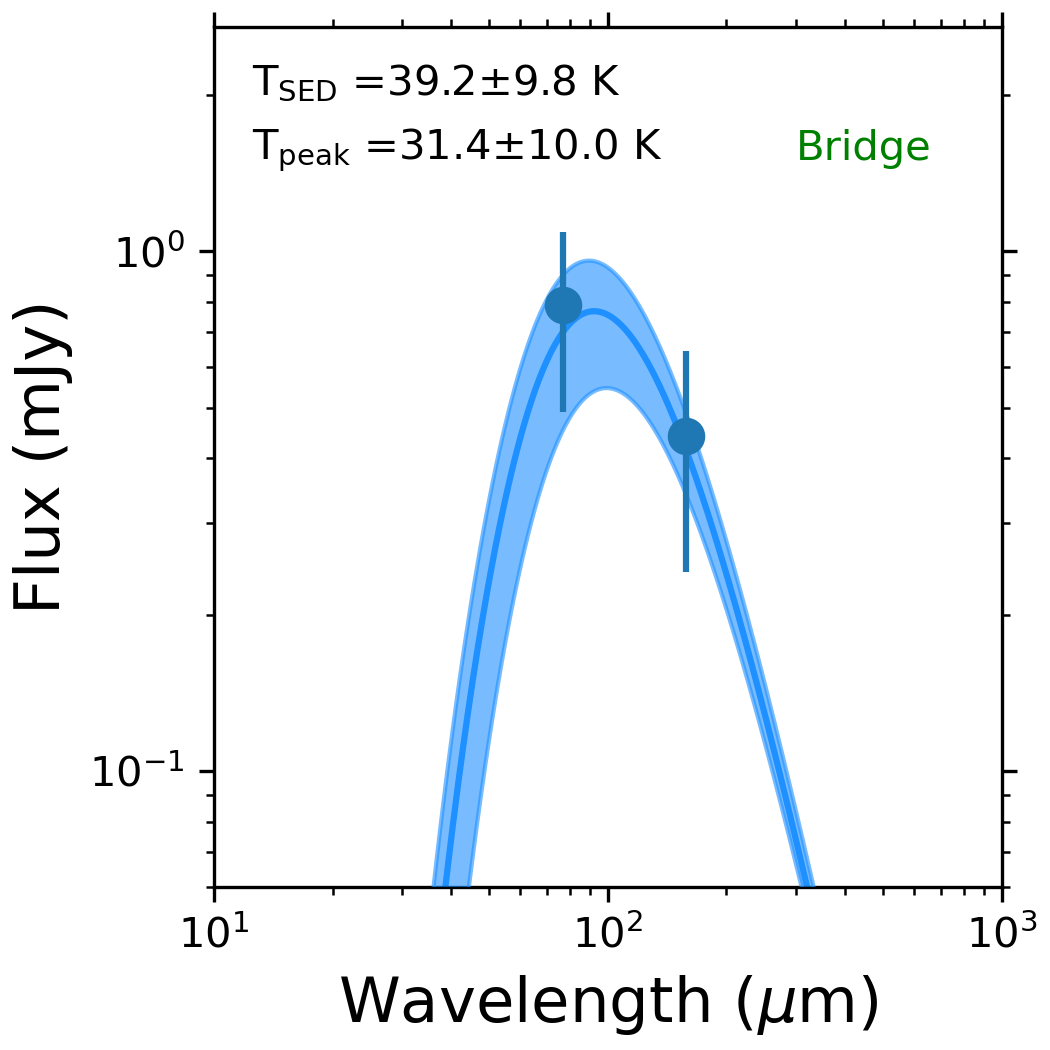} 
  \caption{Best-fittings for far-IR SEDs for 77 and 158 $\mu$m continuum emission, using equations \ref{eq_1}-\ref{eq_3} and adopting the $\beta_{\rm d}$ obtained in Fig. \ref{fig_3}, for the three components identified in the analysis from \S\ref{HZ10_structure}: HZ10-C (left panel), HZ10-W (middle panel), and the Bridge (right panel). Conventions are as in Fig. \ref{fig_3}. %The red-shaded area correspond to the ALMA Band 10 range, which allow us to predict the expected dust continuum fluxes for future observations. 
  \noindent As shown in Figs. \ref{fig_1} and \ref{fig_2}, HZ10-C and the Bridge have lower 77 $\mu$m continuum emission compared to that for HZ10-W (although still within the uncertainties); this may be reflecting a strong UV dust absorption in HZ10-W, which translates into a higher $T_{\rm SED}$ temperature when compared to those of the other two components (although all of them with the same temperature within 1$\sigma$).}
  \label{fig_4}
\end{figure*}

Interestingly, the middle row of Figure \ref{fig_2}\, shows that the HZ10 system cannot be described just considering two S\'ersic profiles. A visual inspection of the residual map of the 158 $\mu$m continuum emission reveals a third component connecting HZ10-C with HZ10-W. Applying the procedure described in \S\ref{Parametric_morphology} to the residual 158 $\mu$m map, we note that the extra component (or the Bridge) cannot be well described by a S\'ersic profile (after trying to fit a S\'ersic profile to the residuals).
%Based on the visual inspection of the residual 158 $\mu$m map (the procedure described in \S\ref{Parametric_morphology}), shown in the right panel of Figure \ref{fig_2}, we note that the extra component (or Bridge) can not be well described by a S\'ersic profile. 
\noindent To analyze the physical properties of the dust at the Bridge, we derive the coordinates of the centroid of its 158 $\mu$m continuum emission (R.A.$_{\rm Int}$, Dec.$_{\rm Int}$); we obtain (10$^{\rm h}$00$^{\rm m}$59$^{\rm s}$.279$\pm$$0^{\rm s}$.037, +01$^{\circ} {33}' {19}''$.463$\pm$${0}''$.038).

The presence of the Bridge may suggest either the interaction or a tidal tale connecting the two main components of the HZ10 system. We are conducting a more detailed analysis of the [CII] morphology to further probe the interacting nature of the system to be published in a forthcoming paper (Telikova et al. in prep). This third component is also presented in new and deep JWST/NIRSpec IFU observations of the main nebular lines in HZ10 \citep[][]{Jones2024}.

%The interaction is also identified by Jones et al. (in prep.), who even find a third component to the east of HZ10-C (referred as HZ10-E) when analyzing the \oiii \, emission at $\lambda = 5007$$\rm \AA$ from JWST/NIRSpec integral field unit (IFU) data. Performing a detailed analysis on the HZ10 kinematics, Telikova et al. (in prep.) confirm the existence of HZ10-E and suggest the existence of a long tail as part of the \cii\, halo of the system. 

\subsection{SED fitting, $T_{\rm SED}$ and $T_{\rm peak}$}
\label{Dust_temperature}
%\label{SED_fitting}

\begin{table*}
%\hspace{-2.75cm}
\resizebox{1.\linewidth}{!}{ 
\begin{tabular}{ccccccccccc}
\hline
Source & $S_{\rm 77\mu m}$ & $S_{\rm 158\mu m}$ & $T_{\rm SED}$ &  $T_{\rm peak}$  & log[$L_{\rm FIR}$] & log[$L_{\rm IR}$] & log[$L_{\rm [CII]}$] & log[$\Sigma_{\rm FIR}$] & $\beta_{\rm UV}$ & IRX \\
& (mJy) & (mJy) & (K) & (K) & ($L_{\odot}$) & ($L_\odot$) & ($L_\odot$) & ($L_\odot$ kpc$^{-2}$) & \\
 (1) & (2) & (3) & (4) & (5) & (6) & (7) & (8) & (9) & (10) & (11)\\
\hline
%1.91e-4, 9.31e-5,  3.5e-5, 2.82e-5
Global & 3.78$\pm$0.63  & 1.76$\pm$0.51 & 46.7$\pm$6.8 & 35.5$\pm$7.0 & 12.42$\pm$0.12 & 12.60$\pm$0.27 & 10.48$\pm$0.14 & 9.59$\pm$0.14 & -1.59$\pm$0.24 & 2.27$\pm$0.98\\

HZ10-C & 1.04$\pm$0.38 & 0.42$\pm$0.20 & 46.4$\pm$11.7 & 35.3$\pm$11.9 & 11.83$\pm$0.38 & 12.01$\pm$0.32 & 9.89$\pm$0.15 & 9.94$\pm$0.15 & -1.63$\pm$0.03 & 1.71$\pm$1.01\\ 
HZ10-W & 1.06$\pm$0.39  & 0.36$\pm$0.13 & 51.2$\pm$13.1 & 37.8$\pm$13.3 & 11.89$\pm$0.45 & 12.06$\pm$0.33 & 9.65$\pm$0.20 & 9.98$\pm$0.20 & -0.64$\pm$0.14 & 2.71$\pm$1.32\\ 
Bridge & 0.79$\pm0.37$  & 0.44$\pm0.17$ & 39.2$\pm$9.8 & 35.9$\pm$10.0 & 11.71$\pm$0.33 & 11.88$\pm$0.37 & 9.76$\pm$0.42 & 9.80$\pm$0.42 & -1.74$\pm$0.15 & 2.38$\pm$1.59\\ 
\hline\hline
\end{tabular}}
\caption{Main properties of the HZ10 system derived in this work. Column (1): component name. Column (2) and (3): 77 and 158 $\mu$m continuum emission flux, respectively. Column (4) and (5): logarithmic of the SED and peak temperatures, respectively. Column (6): logarithmic of the total FIR luminosity computed in the range of 42.5 and 125.5 $\mu$m. Column (7): logarithmic of the total IR luminosity computed in the range of 8 and 1000 $\mu$m. Column (8): total luminosity of the \cii \, emission line computed by using Equation \ref{eq_6}. Column (9): FIR luminosity surface density computed after dividing column (7) by the area of a circular region with a diameter equivalent to the major axis of the ALMA 77 $\mu$m continuum beamsize (i.e., $\sim2.5$ kpc). Column (10): UV spectral slope. Column (11): IR excess.}
\label{table_3}
\end{table*}

We perform far-IR SED fitting in order to determine the physical conditions of the dust in HZ10. We use Equation \ref{eq_0} to compute the global values of the 77 $\mu$m, 110 $\mu$m, 158 $\mu$m, and 198 $\mu$m dust-continuum fluxes, which correspond to the sum of all the flux within a circular area with a diameter equivalent to the major axis of the 198 $\mu$m continuum beamsize. We recall that the latter is the coarsest angular resolution among all the dataset (i.e., $={1.}''2\approx 7$ kpc at HZ10's distance). The aperture is placed at the 158 $\mu$m continuum emission centroid of the Bridge (see \S\ref{HZ10_structure}). To do this, we convolve all the maps to the coarsest angular resolution among the continuum maps (i.e., 198 $\mu$m dust-continuum beamsize). We then use these fluxes to look for the best-fit parameters of equations \ref{eq_1} and \ref{eq_3} by performing the task {\tt minimize} from the {\tt PYTHON} package {\tt lmfit} \citep[][]{Newville2016}. We adopt the {\tt emcee} method, which determines for the maximum likelihood of the parameters via a Monte-Carlo Markov Chain. We finally choose fixed values for $\alpha=2.0$ and $\lambda_{\rm 0}=100$ $\mu$m to perform the fitting (see \S\ref{Basic_equations}). The results, including the best parameters for the far-IR SED fitting and the $1\sigma$ uncertainties curves, are shown in Figure \ref{fig_3}. We highlight that ALMA Band 9 data, which is closer to the SED peak of HZ10 (see Fig. \ref{fig_4}), allow us to reduce the uncertainties by almost three times of $T_{\rm SED}$ estimations when compared to global estimations included in previous studies (e.g.,  \citealt{Faisst2020,Mitsuhashi2023}).

We also perform the far-IR SED fitting for the three sources described in \S\ref{HZ10_structure}. Similarly to way than for the global SED fitting described above, we use a circular aperture with a diameter equivalent to the major axis of the 77 $\mu$m continuum (i.e., the coarsest angular resolution among the \cii\, line, 77, and 158 $\mu$m continuum emission maps; $\sim{0}''.4$, or $\sim2.5$ kpc) to compute the integrated fluxes within such aperture (included in Table \ref{table_3}). The aperture is located at three different positions given by the centers of the 77 $\mu$m continuum S\'ersic profiles for HZ10-C and HZ10-W (see Table \ref{table_2}), and the 158 $\mu$m continuum emission centroid of the Bridge. The integrated fluxes are shown in Table \ref{table_3}.

When comparing our results from the integrated SED fitting with previous studies, we find a dust emissivity index $\beta_{\rm d}=2.00\pm0.14$ slightly lower (although still consistent) than that derived from \cite{Faisst2020} ($\beta_{\rm d}=2.15^{+0.41}_{-0.54}$; based on the SED modelling of Band 6, 7 and 8 data). Conversely, our estimated global SED dust temperature is statistically identical to the one derived by \cite{Faisst2020} to ours ($46.2^{+16.2}_{-8.5}$ and $46.7\pm6.8$ K, respectively). We remark that $\beta_{\rm d}$, $T_{\rm SED}$, and $T_{\rm peak}$ (derived from Equation \ref{eq_4}) are very sensitive to both the assumptions for the fixed values and the completeness of the dust-continuum dataset. %In consequence, their ultimate values will depend on the final shape of the resulting modified SED fitting. 
%\noindent Hereafter in our analysis, we adopt a fixed value for $\beta_{\rm d}=1.9$ (i.e., the value derived from the global SED fitting).

%We then perform the far-IR SED fitting for the three sources described in \S\ref{HZ10_structure}. In a similar way than for the global SED fitting described above, we use a circular aperture with a diameter equivalent to the major axis of the 77 $\mu$m continuum (i.e., the coarsest angular resolution among the \cii\, line, 77, and 158 $\mu$m continuum emission maps; $\sim{0}''.4$, or $\sim3$ kpc) to compute the integrated fluxes within such aperture (included in Table \ref{table_3}). The aperture is located at three different positions given by the centers of the 77 $\mu$m continuum S\'ersic profiles for HZ10-C and HZ10-W (see Table \ref{table_2}), and the 158 $\mu$m continuum emission centroid of the Bridge. The integrated fluxes are shown in Table \ref{table_3}.

For the resolved SED far-IR SED fitting, and considering the caveat above, we adopt a fixed value for $\beta_{\rm d}=2.0$ (i.e., the value derived from the global SED fitting). We find slightly dissimilar $T_{\rm dust}$ estimations when compared to the integrated quantities included in \cite{Faisst2020}. Our results of the far-IR SED fitting for the three components, including the best parameters and the $1\sigma$ uncertainty curves, are shown in Figure \ref{fig_4}. Although still within the uncertainties, we note that the resolved structure of the dust revealed by the data presented in this work shows that the $T_{\rm SED}$ depends critically on the component of the HZ10 system.

While the SED dust temperature in HZ10-C is consistent with integrated values presented by previous studies (46.4 K), we note that HZ10-W's $T_{\rm SED}$  is found to be $\sim5$ K higher, although statistically insignificant (less than 1$\sigma$), than that of the central component and previous global estimations by \cite{Faisst2020} (51.2 and 46.2 K, respectively). Interestingly, we also find that the Bridge has a lower $T_{\rm SED}$ values (although still consistent; 39.2$\pm$9.8) compared to HZ10-C and HZ10-W. The latter seems to be reflecting the detached nature of the Bridge respect to the two main components (perhaps due to outflows) and without close sources of hard radiation fields required to increase its $T_{\rm dust}$ up to temperatures similar to those of HZ10-C and HZ10-W.

\begin{figure*}
    %\hspace{-0.5cm}
  \includegraphics[width=18.2cm]{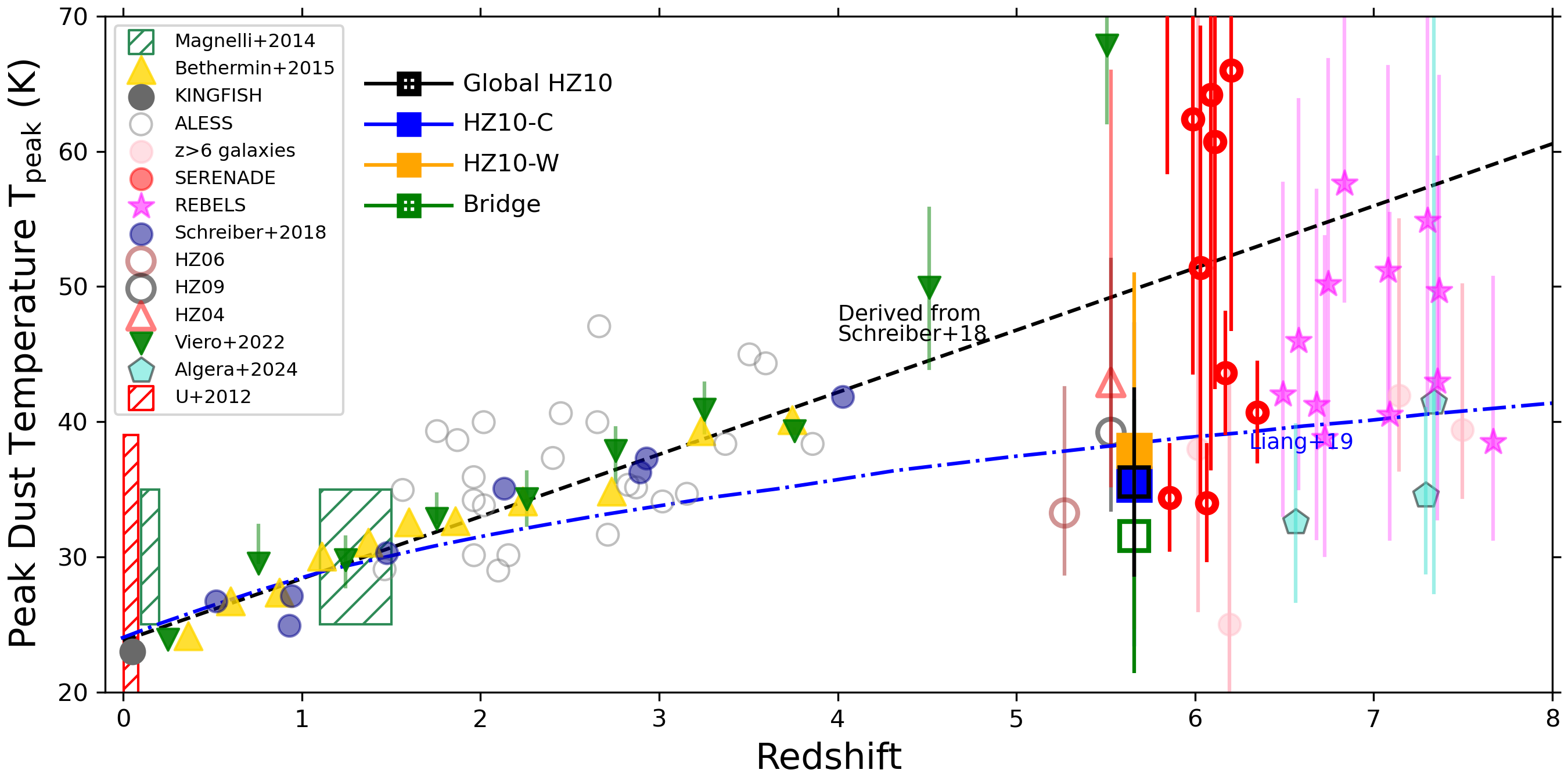} 
  \vspace{-0.25cm}
  \caption{Peak dust temperature ($T_{\rm peak}$) evolution with redshift for HZ10-C, HZ10-W, and the Bridge as blue, orange, and green squares, respectively. The figure also includes galaxy samples at different redshift ranges: $z=0.2-4$ (gray unfilled circles from ALESS; \citealt{daCunha2015}; yellow triangles, \citealt{Bethermin2015}; the two latter samples are shown as included in Fig. 6 from \citealt{Faisst2020}), and $z>6$ (purple circles; \citealt{Knudsen2016,Hashimoto2019}; turquoise pentagons, \citealt{Algera2024}). The red triangle, black circle, and brown circle are galaxies at $z\sim5$ included in \cite{Faisst2020}. Magenta stars correspond to $z\sim 7$ galaxies selected from the Reionization Era Bright Emission Line Survey, REBELS \citep[][]{Sommovigo2022}. Blue large dark-blue circles correspond to galaxies from at $0.5<z<4$ selected from the deep CANDELS fields, and the dashed-black line correspond to their linear best-fit, both extracted from \cite{Schreiber2018}. Green inverted triangles correspond to results based on the stacking analysis and the expected redshift evolution from \cite{Viero2022}. The blue dot-dashed line is the expected peak temperature evolution from hydrodynamic simulation (\citealt{Liang2019,Ma2019}). Red unfilled circles correspond to galaxies at $z\gtrsim6$ selected from the Systematic Exploration in the Reionization Epoch using Nebular And Dust Emission (SERENADE; Harikane et al. in prep.), as included in \cite{Mitsuhashi2023a}. The sample also contains three galaxies selected from \cite{Harikane2020a}. Finally, red and green dashed areas are correspond to the parameter space covered by galaxies selected at different redshfits from the Herschel Multi-tiered Extragalactic Survey (HerMES; \citealt{Magnelli2014}) and ultraluminous infrared galaxies selected from the Great Observatories All-sky LIRG Survey (GOALS; \citealt{U2012}), respectively.}
  \label{fig_5}
\end{figure*}

To put the results from HZ10 in a more general context, Figure \ref{fig_5} shows the evolution of peak dust temperature in star-forming galaxies (expected evolution of the peak temperature derived from Eq. \ref{eq_4}, $T_{\rm peak}$) as a function of redshift. The combination of $T_{\rm peak}$ measurements for low-$z$ galaxies and the interpolation of these estimations at high redshift suggests that temperatures increase up to $z\sim4$ and then flatten off following hydrodynamical simulations from \cite{Liang2019} (blue dashed-dotted line in Fig. \ref{fig_5}). The scatter from the linear relation expected for galaxies at $0.5<z<4$ (see black-dashed line in Fig. \ref{fig_5}) has been shown to depend on many factors. Some of them have been identified in galaxies from the local Universe (e.g., KINGFISH; \citealt{Skibba2011}, GOALS; \citealt{U2012}), including changes in the dust mass density, effects on the dust opacity, and/or variability in the UV luminosity of a central source (e.g., young stars or AGN activity). For instance, \cite{Faisst2017} propose that metallicity also has a significant effect on the $T_{\rm peak}$ measured. In particular, their results indicate that galaxies with faint IR emission and low metallicity could have $T_{\rm peak}$ values similar to those for high IR luminous galaxies. In addition, metallicity can potentially alter the dust properties (e.g., \citealt{Pak1998,Misselt1999,Sommovigo2022}), mainly produce environments with lower opacities (e.g., \citealt{Issa1990,Lisenfeld1998}) due to harder stellar radiation fields are expected for low metallicity environments altering the dust temperature.

Based on the HST WFC3/F160W data of HZ10-W shown in panel A of Figure \ref{fig_1}, the higher $T_{\rm SED}$ may be responding to UV emission severely attenuated by dust, producing a significant increase of the dust temperature. Spectroscopic observations of the main nebular lines with JWST/NIRSpec (e.g., \citealt{Jones2024}) and future ALMA Band 10 continuum and JWST observations could help us to get better constraints on the dust emission and metallicity, allowing us to break down the potential degeneracy of the $T_{\rm SED}$ estimations.

\subsection{The Physical Properties of the ISM in HZ10}
\label{physical_conditions}

\begin{figure}
    %\hspace{-0.5cm}
  \includegraphics[width=9.cm]{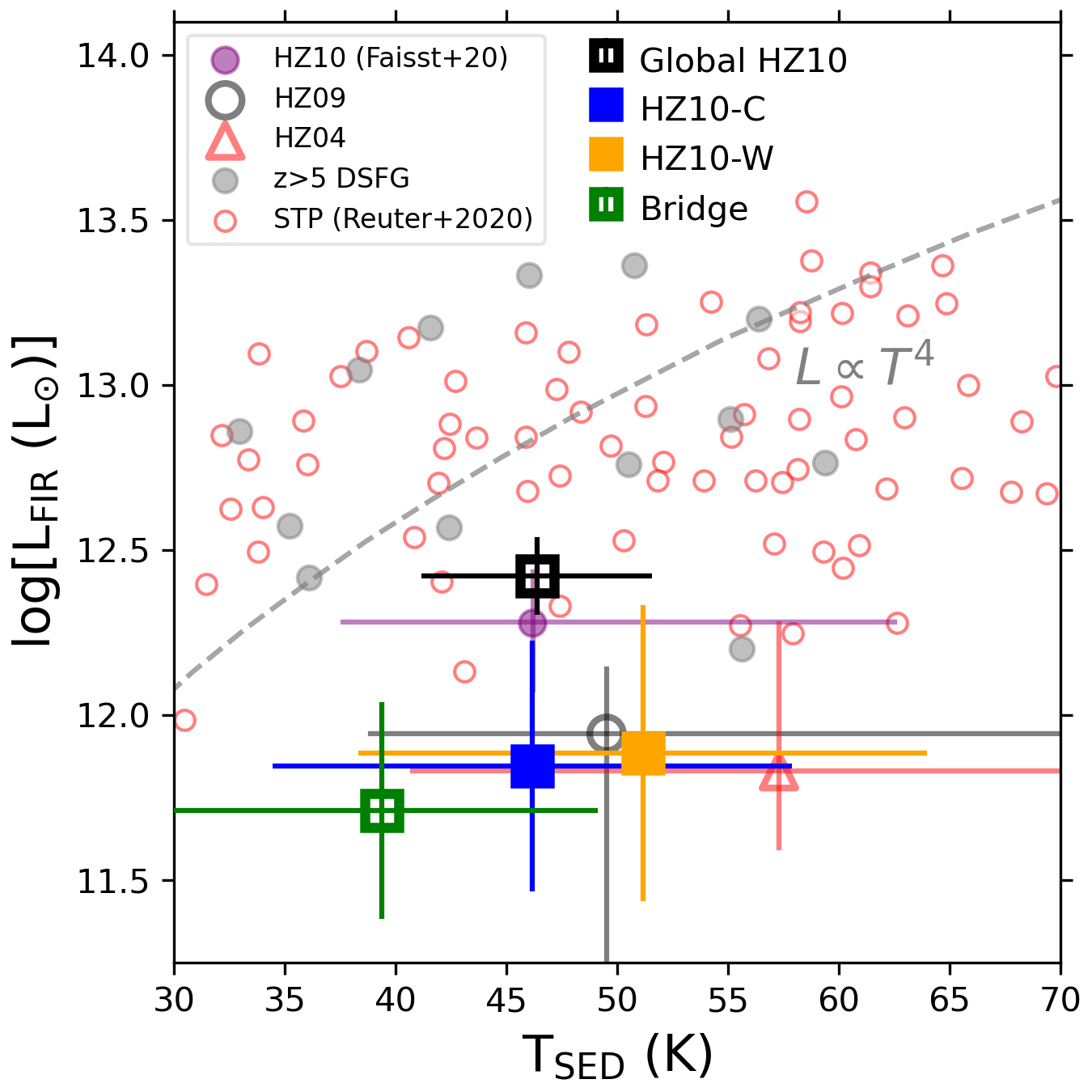}
  \vspace{-0.25cm}
  \caption{Comparison of far-IR luminosity ($L_{\rm FIR}$) and SED temperatures ($T_{\rm SED}$) for the three components of HZ10 covered in this work. The figure includes a sample of dust star-forming galaxies at $z>5$ (gray solid circles; \citealt{Riechers2020}), strongly gravitational lensed dusty star-forming galaxies at $1.9<z<6.9$ included in \cite{Reuter2020}, and previous comparisons for HZ09 (black unfilled circle), HZ04 (red empty triangle), and HZ10 (purple filled circle), as included in \cite{Faisst2020}. The gray dashed line is the $L \propto T^{4}$ relation, also extracted from \cite{Faisst2020}.}
  \label{fig_6}
\end{figure}

Figure \ref{fig_6} presents the relation between far-IR luminosity, $L_{\rm FIR}$, and $T_{\rm SED}$ for the three components of HZ10 identified in this work. The values of $L_{\rm FIR}$ for HZ10-C, HZ10-W, and the Bridge are consistent with the global far-IR luminosity derived by \cite{Faisst2020}. However, compared to the expected relation from the optically thick case (i.e., where $L\propto T^{4}$; grey-dashed line in Fig. \ref{fig_6}), the HZ10 components seem to be either too warm (i.e., to the right of the relation) or underluminous in the far-IR (thus, below the relation). In addition, when compared to dusty star-forming galaxies at similar redshifts (DSFG, included as gray solid circles in Fig. \ref{fig_6}; \citealt{Riechers2020}), our $L_{\rm FIR}$ values may be reflecting intrinsic differences of the dust properties between some CRISTAL and DSFG systems, but most likely due different spatial configuration and/or optically thickness. Some of these differences could respond as well to the factors described at the end of \S\ref{Dust_temperature}, including the effects of metallicity, dust abundance on the dust opacity \citep{Faisst2017}, or changes in the photoelectric efficiency of the dust (e.g., \citealt{Nath1999,Malhotra2017,McKinney2021,Glatzle2022}).

%Are the physical properties of the dust impacting the star-formation activity in HZ10? 
The \cii-to-FIR luminosity ratio, $\cii/$FIR=$L_{\rm [CII]} / L_{\rm FIR}$, has been found to be closely related to the physical properties of the ISM. For instance, while in low-metallicity environments the typical values are log($\cii/$FIR)$\sim-2$ (e.g., dwarf galaxies or star-forming disks; \citealt{Madden2013,Smith2017,Herrera-Camus2018a}), in nuclear regions, starburst systems, and AGNs are around log($\cii/$FIR)$\sim$[-4,-3] (e.g., \citealt{Malhotra2001,DiazSantos2013,Herrera-Camus2018a}).

%; a decrease could potentially be linked to the gas ionization state due to AGN activity (e.g., \citealt{Herrera-Camus2018b,Langer&Pineda2015}), an increase in the ionization parameter of the \hii\, regions (e.g., \citealt{Gracia-Carpio2011,Herrera-Camus2018b}), among others.

\begin{figure}
    %\hspace{-0.5cm}
  \includegraphics[width=9.cm]{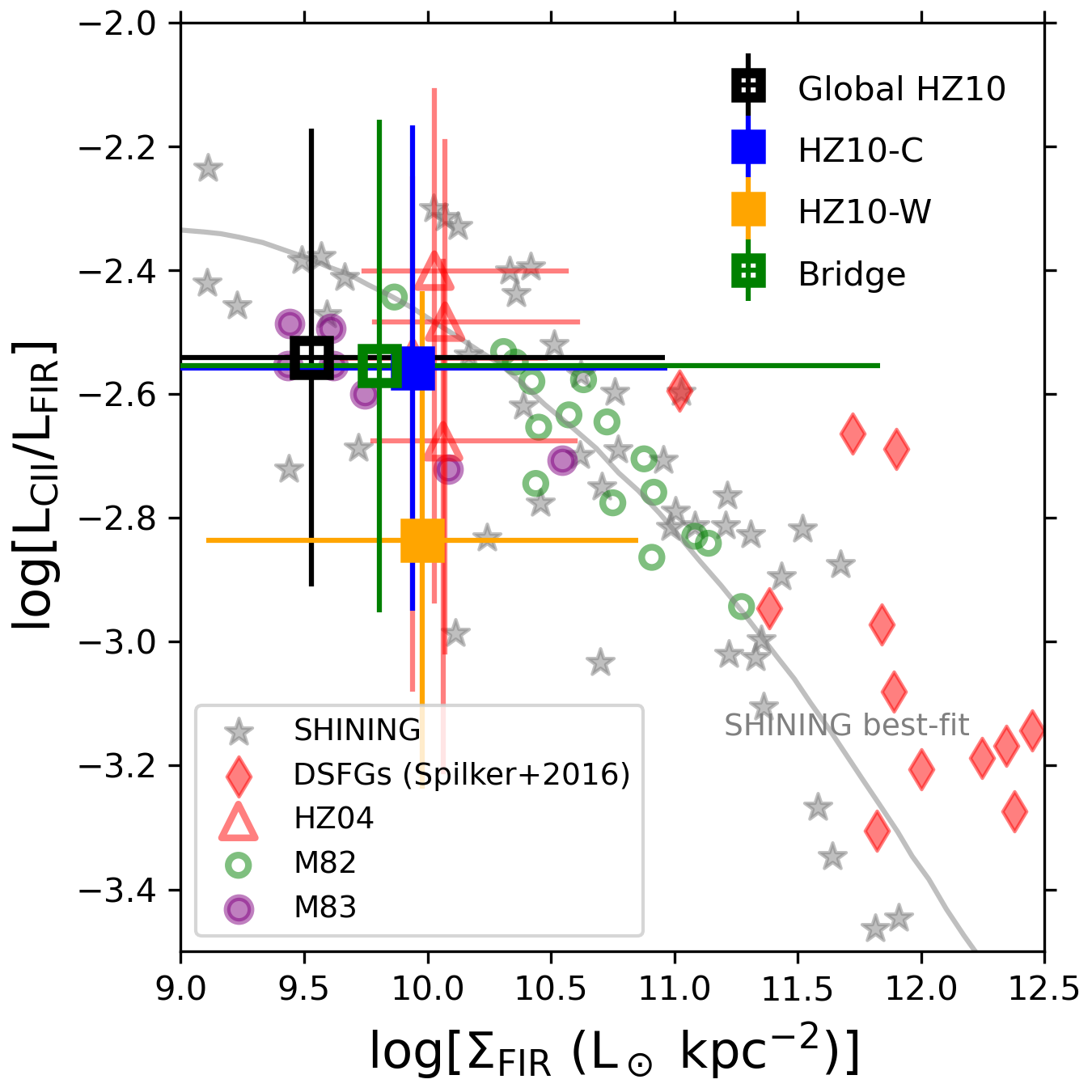} 
  \caption{\cii /FIR ratio as a function of the FIR surface density ($\Sigma_{\rm FIR}$) for HZ10-C (blue square), HZ10-W (orange square), and the Bridge between them (green empty square). For comparison, the figure also includes nearby star-forming and starburst galaxies from the SHINING sample (grey filled stars; \citealt{Herrera-Camus2018a}), $\sim 100$ pc scale regions in the central disk of M82 (green unfilled dots; \citealt{Contursi2013,Herrera-Camus2018a}), $\sim 400$ pc scale regions from central regions of M83 (purple filled circles), lensed dusty star-forming galaxies at $z\sim 1.9-5.7$ (DSFGs; red-solid diamonds; \citealt{Spilker2016}) , and four kpc-size regions extracted across the disk of HZ04 (red unfilled triangles; \citealt{Herrera-Camus2021}). The The solid grey line corresponds to the best quadratic fit to the SHINING data (as included in \citealt{Herrera-Camus2018a}).}
  \label{fig_7}
\end{figure}

Figure \ref{fig_7} shows $\cii/$FIR as a function of the far-IR luminosity surface density, $\Sigma_{\rm FIR}$, for the three components of HZ10 analyzed in this work. We note that HZ10-C, HZ10-W, and the Bridge have a very smooth distribution of \cii-to-FIR values, with a variation at most as $25$\% between the faintest (i.e., the Bridge) and the strongest continuum emission (i.e., HZ10-W). Although differences in the \cii\, emission are slightly more significant, these are at 35-45\% between the faintest (i.e., the Bridge and HZ10-W) and the strongest \cii\, emitter (HZ10-C). Our results are consistent with those of $\sim100$ pc scale regions in the central disk of the nearby starburst galaxy M82 (green unfilled dots; \citealt{Contursi2013,Herrera-Camus2018a}). Following a similar methodology to that described in \S\ref{Dust_temperature} to the physical parameters of the dust in a typical star-forming galaxy at $z\sim5.5$ ($\beta_{\rm d}=1.5$, $T_{\rm SED}=45$ K; e.g., \citealt{Pavesi2016,Faisst2017}), \cite{Herrera-Camus2021} compute the $\cii/$FIR-$\Sigma_{\rm FIR}$ relation for HZ04 (CRISTAL-20 in the CRISTAL survey). Splitting the analysis in four independent beams (red unfilled triangles in Fig. \ref{fig_7}), they find log($\cii/$FIR)$\sim$$[-2.7,-2.3]$ and log($\Sigma_{\rm FIR}$/[L$_\odot$ kpc$^{-2}$])$\sim$10. Although $\cii/$FIR values for our sources are comparable to those for HZ04, we note that  our sources have $\Sigma_{\rm FIR}$ values (directly the star-formation rate, SFR) around 4 times higher than those derived in \cite{Herrera-Camus2021}. These results may suggest a deficit in the \cii\, content in the HZ10 system, specially in HZ10-W, although we note that the dust SED setup used in \cite{Herrera-Camus2021} is different to ours and therefore this may affect the comparison between the two sources given its impact on the derived IR luminosities.

Several studies have shown the strong impact that AGNs could potentially have on the \cii\, line emission, producing low \cii-to-FIR luminosity ratios in nearby galaxies (e.g., \citealt{Stacey2010,Sargsyan2012,Herrera-Camus2018b}). \cite{DiazSantos2013} suggest that in galaxies with a strong \cii\, deficit (log($\cii/$FIR)$<-3$) the AGNs can play an important role in destroying significant fractions of polycyclic aromatic hydrocarbon (PAH) molecules (i.e., associated with \cii\, emission) due to their hard radiation field (e.g., \citealt{Lai2023}). This is consistent with the evidence (although marginal) from the new JWST/NIRSpec observations that suggest that HZ10-W may contain nuclear activity \citep[][]{Jones2024}.
%proposed two cases to explain the \cii\, deficit depending on the $\cii/$FIR value. They suggest that while for sources with log($\cii/$FIR)$<-3$ (i.e., extremely low values), AGNs can play an important role in destroying significant fractions of polycyclic aromatic hydrocarbon (PAH) molecules (i.e., associated with \cii\, emission) due to their hard radiation field, for log($\cii/$FIR)=[-2,-1] the \cii\, deficit is likely to respond to properties of the starburst region itself instead. 
%Although Jones et al. (in prep.) suggest that HZ10-W may contain nuclear activity, they only find poor evidence for high-ionization regions based on the analysis of the optical lines emission available (E.G., CIV $\lambda=$1548,1551, and NIV$\lambda=$1719). 
On the other hand, by modeling the emission of the molecular, neutral, and ionized gas of galaxies selected from the Survey with {\it Herschel} of the Interstellar medium in INfrared Galaxies (SHINING), \cite{Gracia-Carpio2011} found that a decrease in $\cii/$FIR ratios can be explained by increasing the value of the ionization parameter ($U$) on the surface of molecular clouds. As $U$ increases, a larger fraction of UV photons are absorbed by dust in the ionized region and reemitted in the form of infrared emission. The net effect is that the fraction of UV photons available to ionize and excite the gas is reduced at high $U$, decreasing the relative intensity of the fine structure lines compared to the FIR continuum (e.g, \citealp{Voit1992,Abel2009,Gracia-Carpio2011,Herrera-Camus2018a}). However, the UV photons could be also interacting with gas itself, having a potential impact on its physical properties (i.e., the \cii\,line emission).
%Since we have shown that HZ10-W has a $\cii/$FIR value very close to the former limit, AGN activity may also be a plausible mechanism to explain its slight \cii\, deficit. 
%\noindent {\color{red} Future studies based on optical IFU data (e.g., Jones et al. in prep.), will study in detailed the feasibility of the different scenarios (using metallicity prescriptions) taking place in HZ10-W.}

\subsection{The IRX-$\beta_{\rm UV}$ relation}
\label{IRX-beta}

Unbiased galaxy star-formation rates at low and high redshift are essential to get a complete picture of the mechanisms changing the physical properties of the ISM as a function of cosmic times. It is critical thus to account for both the dust thermal emission and the UV light to mitigate the natural biases on the derivation of accurate SFR estimations. In this sense, the IRX-$\beta_{\rm UV}$ relation (e.g., \citealt{Meurer&Heckman1995,Meurer1999}) has shown to be a strong observational tool to link the IR emission to UV measurements, particularly to the tight correlations of the IRX-$\beta_{\rm UV}$ plane exhibited at different redshift ranges (e.g., \citealt{Heinis2013,McLure2018,Fudamoto2020,Bowler2024}). In the last decades, however, several studies have revealed that some local LIRGs/ULIRGs (e.g., \citealt{Howell2010}) and high-$z$ galaxies (e.g., \citealt{Alvarez-Marquez2016, Bouwens2016, Reddy2018}) can depart from this relation due to changes in intrinsic dust properties. Such variations can be produced by the composition of the dust, the spatial distribution of the dust and UV emission, or due to ISM turbulence, among others (see \citealt{Liang2021} and references therein). 

\begin{figure}
    %\hspace{-0.5cm}
  \includegraphics[width=9.cm]{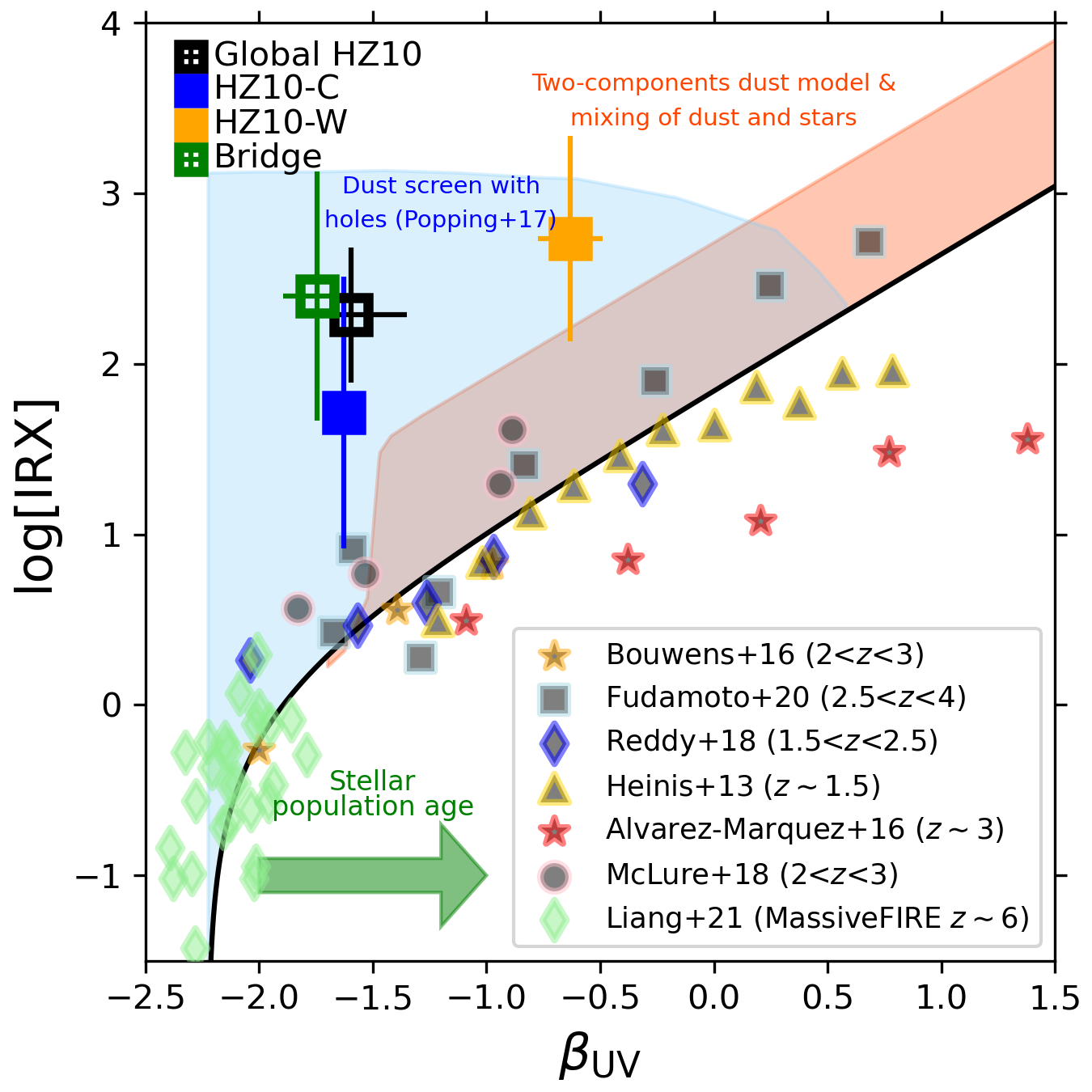} 
  \caption{Comparison between the IR excess (IRX) and the UV spectral slope ($\beta_{\rm UV}$); i.e., the IRX-$\beta_{\rm UV}$ relation for HZ10-C, HZ10-W, and the Bridge between them. Conventions are as in Fig. \ref{fig_7}. Green diamonds correspond to the {\tt MASSIVE}FIRE sample at $z=6$, as included in \citealt{Liang2021}. The figure also encompasses a series of galaxy samples from the literature: yellow triangles are taken from \cite{Heinis2013}, red stars from \cite{Alvarez-Marquez2016}, orange stars from \cite{Bouwens2016}, blue-edge diamonds from \cite{Reddy2018}, magenta-edged circles from \cite{McLure2018}, and cyan-edged squares from \cite{Fudamoto2020}. The black solid line is the best linear-fit for the IRX-$\beta_{\rm UV}$ relation for UV-selected starburst galaxies as shown in \cite{Meurer1999}. The light-blue/light-orange shaded areas, and the green arrow at the bottom are extracted from the schematic figure included in \cite{Popping2017}, which summarizes the different physical mechanisms affecting the properties of the emission sources.}
  \label{fig_8}
\end{figure}

To test which is the more adequate scenario in the HZ10 system, we use equations \ref{eq_7} and \ref{eq_8} to drawn the IRX-$\beta_{\rm UV}$ relation for HZ10-C, HZ10-W, and the Bridge, as shown in Figure \ref{fig_8}. In addition to several galaxy samples at different redshifts, we have also overplotted some of the schematics from Figure 11 in \cite{Popping2017}, which reflect the distinct physical mechanisms affecting the properties of the emission sources (see light-blue and light-orange shaded areas, and the green arrow at the bottom of Fig. \ref{fig_8}). While HZ10-C is close to the best linear-fit for UV-selected near starburst galaxies (black solid line; \citealt{Meurer1999}), we note that both HZ10-W and the Bridge lie significantly above it; the former also has significantly a higher $\beta_{\rm UV}$ value compared to that for the other two members ($\beta_{\rm UV}\approx -0.64$), which is consistent with the results included in \cite{Jones2024} . According to \cite{Popping2017} models, the location of HZ10-W and the Bridge in the IRX-$\beta_{\rm UV}$ relation seems to favor scenarios where a screen of dust is placed with holes in between a relative young stellar population and the observer. Although a small level of turbulence within the dust could also increase its optical depth (e.g. \citealt{Fischera2003,Popping2017}), the dust screen appears to be a simple (yet reasonable) explanation for the dust emission in the more dusty components of the HZ10 system. For example, although with higher $\beta_{\rm UV}$ than those for the three components analyzed in this work, the star-forming main-sequence galaxies analyzed by \cite{Fudamoto2020} have similar IRX values than HZ10's. They propose that the high dust-attenuation properties shown by those galaxies may correspond to supernovae (SNe) driven dust production at $z\geq2-3$. Such SNe dust could be consistent with the steeper dust curve observed at $z\sim 4-6$ galaxy sample compared to the attenuation inferred at $z<3$ for sources similar to those included in \cite{Meurer1999}(e.g., \citealt{Maiolino2004,Hirashita2005,Gallerani2010}).

Numerical simulations have also found that \cii/FIR ratios can decrease with increasing $\Sigma_{\rm FIR}$, leading to an apparent \cii\, deficit. In particular, \cite{Bisbas2022} show that this could reflect the thermal saturation of \cii\, as a consequence of the strong far-UV heating related to the high SFR. They propose that while the \cii\, emissivity increase asymptotically in this regime, the FIR emission increases linearly, leading the deficit of the \cii. 

The IRX-$\beta_{\rm UV}$ relation has been shown to be a powerful tool to characterize the complex structure of the HZ10 system. However, some studies have suggested the that IRX-$\beta_{\rm UV}$ relation may fail in probing the physical conditions of the ISM in galaxies with large IR-to-UV flux ratios compared to the $\beta_{\rm UV}$ (e.g., \citealt{Ferrara2022}). To address this problem, \cite{Ferrara2022} introduce the non-dimensional ``molecular index'', $I_{\rm m}=(F_{158\mu m}/F_{\rm 1500\AA} )/(\beta_{\rm UV}-\beta_{\rm int})$; here, $F_{158\mu m}$ and $F_{\rm 1500\AA}$ are the fluxes at $1500\AA$ and the observed far-infrared continuum flux at 158 $\mu$m, respectively, and $\beta_{\rm int}$ is the intrinsic UV spectral slope (typically, $\beta_{\rm int}\approx -2.406$; see \citealt{Ferrara2022} for more details). When applying to sources selected from REBELS, they note that $I_{\rm m}$ is a good predictor of the simultaneous presence of optically thin and thick regimes, showing that galaxies with a two-phase medium have $I_{\rm m}>1120$. When computing the molecular index for the HZ10 system, we obtain $I_{\rm m} = 1156\pm104$. The latter is consistent with the IRX-$\beta_{\rm UV}$ relation for HZ10 discussed above, supporting the scenario of a two-phase structure with young stars embedded in optically thick giant molecular clouds and an older stellar populations immersed in a more transparent medium.

Future ALMA-CRISTAL studies (Killi et al. in prep.) will analyze the variations of the IRX-$\beta_{\rm UV}$ relation among CRISTAL galaxies on $\sim$kiloparsec scales, therefore allowing us to derive more statistically significant results and to perform a meticulous comparison with other similar studies from the literature.

%\subsection{A final note on future observations}
%\label{ALMA_BAND_10}

%As mentioned previously, $T_{\rm SED}$ estimations (and related quantities) are very sensitive to the completeness of the integrated FIR continuum fluxes set available to model the dust SED. Complementary continuum fluxes measurements at higher frequencies than those covered by ALMA Band 9 data could thus allow us to get better constraints on the dust temperature of the different components in HZ10. In particular, it is necessary to check possibles scenarios where the dust peak is located at higher frequencies than those covered by Band 9. ALMA Band 10, centered at $\lambda_{\rm Band10}=$350 $\mu$m, is a good candidate to address this issue for $z\sim4-6$ galaxies. Since it gives spectral coverage in the wavelength range $\sim [45,60]$ $\mu$m at the redshift of HZ10, ALMA Band 10 provides a valuable option for a better characterization of dust continuum emission.

%The uncertainties of the SEDs included in Figure \ref{fig_4} for HZ10-W (which has a particularly higher $T_{\rm SED}$ compared to previous estimations; see \S \ref{Dust_temperature}), allows us to predict the expected lower-limits of the dust continuum fluxes within the ALMA Band 10 coverage for data at $\sim {0}''.3$. According to our calculations, we expect a lower-limit flux of $F_{\rm Band10}\sim 0.2$ mJy, which translates in a required sensitivity $\sim 20$ $\mu$Jy beam$^{-1}$ (adopting a Gaussian beamsize of axes $\approx {0}''.3$). Based on this sensitivity, the ALMA Sensitivity Calculator (ASC\footnote{https://almascience.eso.org/proposing/sensitivity-calculator}) gives us an integration time of $\approx 1.6$ h. This reasonable amount of time required to reach the mentioned sensitivity will allow us to probe HZ10-W dust temperatures as low as $\sim 46$ K, $T_{\rm SED}$ similar or higher than the one derived in this work.%, and/or rule out scenarios with $T_{\rm SED} \gtrsim 60$ K.

%However, future ALMA Band 10 continuum %and JWST 
%\noindent data could help us to get better constraints on the dust emission and metallicity, therefore allowing us to break down the potential degeneracy of the $T_{\rm SED}$ estimations.

\section{Summary and conclusions}
\label{S5_Conclusions}

We present a study of the dusty, main-sequence galaxy HZ10 at $z\approx 5.66$, as part of the CRISTAL survey. We present new ALMA \cii\, line emission and Band 9 dust continuum data, which is key to constrain the peak of the dust SED. In combination with ALMA Band 6, 7, 8 and HST WFC3 archival data, we conduct a systematic analysis of the morphology and the physical conditions of dust and gas in such multi-component system. We characterize the main properties of the structures identified in HZ10, such as the SED and peak dust temperatures, FIR luminosities, \cii/FIR ratios, UV spectral slopes, and their IR excesses, and we compare our results with the current literature. Our main conclusions are enumerated as follows:

\begin{enumerate}
    \item We perform a parametric 2D fitting to derive the morphological parameters of the two main components of HZ10 (HZ10-C and HZ10-W), which are well described by S\'ersic profiles in the 77 and 158 $\mu$m dust continuum and the \cii\, line emission data (see Table \ref{table_2}). Interestingly, the residual map of the 158 $\mu$m dust continuum data reveals a third component (Bridge), which seems to correspond to a bridge-like dusty structure connecting the central and west components.

    \item We carry out a global modified blackbody SED fitting of HZ10 using the 77, 110, 158, and 198 $\mu$m dust continuum fluxes. We derive a global dust emissivity index $\beta_{\rm d} \approx 2.0\pm0.14$ and a global dust temperature of $T_{\rm SED}=46.7\pm6.8$ K. We adopt the global value of $\beta_{\rm d}$ to perform modified SED fittings of the three main structures identified in the HZ10 system using the 77 and 158 $\mu$m dust continuum maps (since these two datasets allow us to resolve spatially the three components of HZ10), obtaining spatially-resolved estimations of their SED and peak dust temperatures. We find that HZ10-W, which is the component that shows the higher obscuration of the rest-frame UV emission, has a dust temperature of $T_{\rm SED}=51.2\pm13.1$ K, which is about $\sim5$ K higher than the two other components (although all of them with the almost the same SED temperature within 1$\sigma$). More importantly, the inclusion of the new ALMA Band 9 continuum data allow us to reduce the uncertainties in the global dust temperature measurements by a factor of $\sim2.3$. 

    \item We compute the \cii-to-FIR luminosity ratio, \cii/FIR, for HZ10-C, HZ10-W, and the Bridge. When we compare \cii/FIR with the FIR surface density, $\Sigma_{\rm FIR}$, we find that our sources cover a similar parameter space to that of local starburst galaxies. We note that HZ10-W shows signs of a \cii\, deficit, suggesting possibles scenarios such as hard radiation field destroying PAHs associated with \cii\, emission (e.g., young stellar populations or AGN activity), or variations in the dust photoelectric efficiency.

    \item We calculate the IR excesses and the UV spectral slopes (the IRX-$\beta_{\rm UV}$ relation), for the three components in HZ10. While HZ10-C has IRX and $\beta_{\rm UV}$ values consistent with those of UV-selected starburst local galaxies and other high-$z$ galaxies, both HZ10-W and the Bridge clearly depart from the observed sequence of global galaxies the IRX-$\beta_{\rm UV}$ relation. According to theoretical models from previous studies, our results suggest that the UV-emission in HZ10-W and the Bridge may be strongly attenuated by a dust screen in between young stellar populations and the observer.
     
\end{enumerate}

As mentioned previously, $T_{\rm SED}$ estimations (and related quantities) are very sensitive to the completeness of the integrated FIR continuum fluxes set available to model the dust SED. Complementary continuum fluxes measurements at higher frequencies than those covered by ALMA Band 9 data could thus allow us to get better constraints on the dust temperature of the different components in HZ10. In particular, it is necessary to check possibles scenarios where the dust peak is located at higher frequencies than those covered by Band 9. ALMA Band 10, centered at $\lambda_{\rm Band10}=$350 $\mu$m, is a good candidate to address this issue for $z\sim4-6$ galaxies. Since it gives spectral coverage in the wavelength range $\sim [45,60]$ $\mu$m at the redshift of HZ10, ALMA Band 10 provides a valuable option for a better characterization of dust continuum emission.

Upcoming ALMA-CRISTAL studies will analyze the \cii, kinematics and morphologies of HZ10, and the variations of the IRX-$\beta_{\rm UV}$ relation among CRISTAL galaxies in more detail (e.g., Telikova et al. in prep.; Ikeda et al. in prep.; Killi et al. in prep.). In addition, future ALMA Band 10 continuum and JWST (e.g., \citealt{Jones2024}) data will help us to get better constraints on the dust emission and metallicity of HZ10, therefore allowing us to break down the potential degeneracy of our dust SED temperature estimations and related physical quantities. In addition, ALMA band 10 data would allow us to trace the small $T_{\rm SED}$ differences and reducing their uncertainties in the HZ10 system.

%\vspace{1cm}
%\begin{acknowledgments}

\begin{acknowledgements}
V. V. acknowledges support from the ALMA-ANID Postdoctoral Fellowship under the award ASTRO21-0062. R.H.-C. thanks the Max Planck Society for support under the Partner Group project "The Baryon Cycle in Galaxies" between the Max Planck for Extraterrestrial Physics and the Universidad de Concepción. R.H-C. also gratefully acknowledge financial support from ANID BASAL projects FB210003. R. I. is supported by Grants-in-Aid for Japan Society for the Promotion of Science (JSPS) Fellows (KAKENHI Grant Number 23KJ1006). N.M.F.S. acknowledges financial support from the European Research Council (ERC) Advanced Grant under the European Union's Horizon Europe research and innovation programme (grant agreement AdG GALPHYS, No. 101055023). K. T. acknowledges support from JSPS KAKENHI grant No. 23K03466. M. K. was supported by the ANID BASAL project FB210003. K. T. was supported by ALMA ANID grant number 31220026 and by the ANID BASAL project FB210003. M. R. acknowledges support from project PID2020-114414GB-100, financed by MCIN/AEI/10.13039/501100011033. M. S. was financially supported by Becas-ANID scolarship \#21221511, and also acknowledges ANID BASAL project FB210003. M. A. acknowledges support from FONDECYT grant 1211951, and ANID BASAL project FB210003. R. J. A. was supported by FONDECYT grant number 1231718 and by the ANID BASAL project FB210003. R.B. acknowledges support from an STFC Ernest Rutherford Fellowship [grant number ST/T003596/1].
This paper makes use of the following ALMA data: ADS/JAO.ALMA \#2022.1.00678.S, ADS/JAO.ALMA \#2019.1.01075.S, ADS/JAO.ALMA \#2018.1.00348.S, ADS/JAO.ALMA \#2015.1.00388.S. This research is based on observations made with the NASA/ESA Hubble Space Telescope obtained from the Space Telescope Science Institute, which is operated by the Association of Universities for Research in Astronomy, Inc., under NASA contract NAS 5–26555. These observations are associated with program 13641 (P.I.: Peter Capak). The National Radio Astronomy Observatory is a facility of the National Science Foundation operated under cooperative agreement by Associated Universities, Inc.
\end{acknowledgements}

%\end{acknowledgments}

{\it Software: Astropy \citep{AstropyCollaboration2018}, MatPlotLib \citep{Hunter2007},
NumPy \citep{Harris2020}, PYAUTOGALAXY \citep{Nightingale2023},  SciPy \citep{2020SciPy-NMeth}, seaborn \citep{Waskom2021}, Scikit-learn \citep{scikit-learn}.}


\bibliographystyle{aa} % style aa.bst
\bibliography{main}

\end{document}